\documentstyle[a4p,12pt,rotating,subfigure,amsbsy,twoside,epsfig,pennames]{article}

\input epsf % for EPS Figure includes

\def\xE{\ifmmode {\rm x_E} \else {$\rm x_E$} \fi}

\def\costs{\ifmmode {\cos\theta^*} \else $\cos\theta^*$ \fi }

\def\Z0{\ifmmode {{\mathrm Z^0}} \else {${\mathrm Z^0}$}\fi}

\def\ep{\ifmmode {\mathrm{e}^{+}} \else $\mathrm{e}^{+}$ \fi }
\def\epem{\ifmmode {\mathrm{e}^{+}\mathrm{e}^{-}} \else $\mathrm{e}^{+}\mathrm{e}^{-}$\fi }

\def\Dstarpm{\ifmmode {{\mathrm D}^{*\pm}} \else {${\mathrm D}^{*\pm}$}\fi}
\def\Dstarp{\ifmmode {{\mathrm D}^{*+}} \else {${\mathrm D}^{*+}$}\fi}
\def\Dstarm{\ifmmode {{\mathrm D}^{*-}} \else {${\mathrm D}^{*-}$}\fi}
\def\Dstaro{\ifmmode {{\mathrm D}^{*0}} \else {${\mathrm D}^{*0}$}\fi}
\def\Dstar{\ifmmode {{\mathrm D}^*} \else {${\mathrm D}^*$}\fi}
\def\Kstar{\ifmmode {{\mathrm K}^*} \else {${\mathrm K}^*$}\fi}
\def\Kstarp{\ifmmode {{\mathrm K}^{*+}} \else {${\mathrm K}^{*+}$}\fi}
\def\Kstarm{\ifmmode {{\mathrm K}^{*-}} \else {${\mathrm K}^{*-}$}\fi}
\def\Ds{\ifmmode {{\mathrm{D}^{+}_{\mathrm{s}}}} \else {${\mathrm{D}^{+}_{\mathrm{s}}}$}\fi}
\def\ds{\ifmmode {{\mathrm{D}^{+}_{\mathrm{s}}}} \else {${\mathrm{D}^{+}_{\mathrm{s}}}$}\fi}
\def\dplusm{\ifmmode {{\mathrm D}^-} \else {${\mathrm D}^-$}\fi}
\def\dplusp{\ifmmode {{\mathrm D}^+} \else {${\mathrm D}^+$}\fi}
\def\dplus{\ifmmode {{\mathrm D}^+} \else {${\mathrm D}^+$}\fi}
\def\Dplus{\ifmmode {{\mathrm D}^+} \else {${\mathrm D}^+$}\fi}
\def\D0{\ifmmode {{\mathrm D^o}} \else {${\mathrm D^o}$}\fi}
\def\d0{\ifmmode {{\mathrm D^o}} \else {${\mathrm D^o}$}\fi}
\def\pion{\ifmmode {{\mathrm \pi}} \else {${\mathrm \pi}$}\fi}
\def\pion{\ifmmode {{\mathrm \pi_{s}}} \else {${\mathrm \pi_{s}}$}\fi}
\def\K{\ifmmode {{\mathrm K}} \else {${\mathrm K}$} \fi}
\def\g{\ifmmode {{\mathrm g}} \else {${\mathrm g}$} \fi}
\def\u{\ifmmode {{\mathrm u}} \else {${\mathrm u}$} \fi}
\def\d{\ifmmode {{\mathrm d}} \else {${\mathrm d}$} \fi}
\def\s{\ifmmode {{\mathrm s}} \else {${\mathrm s}$} \fi}
\def\c{\ifmmode {{\mathrm c}} \else {${\mathrm c}$} \fi}
\def\b{\ifmmode {{\mathrm b}} \else {${\mathrm b}$} \fi}
\def\q{\ifmmode {{\mathrm q}} \else {${\mathrm q}$} \fi}
\def\uu{\ifmmode {{\mathrm u\bar{\mathrm u}}}
    \else {${\mathrm u\bar{\mathrm u}}$} \fi}
\def\dd{\ifmmode {{\mathrm d\bar{\mathrm d}}}
    \else {${\mathrm d\bar{\mathrm d}}$} \fi}
\def\ss{\ifmmode {{\mathrm u\bar{\mathrm u}}}
    \else {${\mathrm s\bar{\mathrm s}}$} \fi}
\def\cc{\ifmmode {{\mathrm c\bar{\mathrm c}}}
    \else {${\mathrm c\bar{\mathrm c}}$} \fi}
\def\bb{\ifmmode {{\mathrm b\bar{\mathrm b}}}
    \else {${\mathrm b\bar{\mathrm b}}$} \fi}
\def\qq{\ifmmode {{\mathrm q\bar{\mathrm q}}}
    \else {${\mathrm q\bar{\mathrm q}}$} \fi}
\def\e{\ifmmode {{\mathrm e}} \else {${\mathrm e}$} \fi}
\def\p0{ \ifmmode {\pi^{\mathrm{o}}} \else $\pi^{\mathrm{o}}$ \fi }
\def\mus{\ifmmode {\ \mathrm{\mu s}} \else $\mathrm{\mu s}$ \fi}
\def\mum{\ifmmode {\ \mathrm{\mu m}} \else $\mathrm{\mu m}$ \fi}
\def\GeV{\ifmmode { \mathrm{GeV}} \else $\mathrm{GeV}$ \fi}
\def\GeVc{\ifmmode { \mathrm{GeV/c}} \else $\mathrm{GeV/c}$ \fi}
\def\GeVcc{\ifmmode { \mathrm{GeV/c^2}} \else $\mathrm{GeV/c^2}$ \fi}
\def\MeV{\ifmmode {\ \mathrm{MeV}} \else $\mathrm{MeV}$ \fi}
\def\MeVc{\ifmmode {\ \mathrm{MeV/c}} \else $\mathrm{MeV/c}$ \fi}
\def\MeVcc{\ifmmode {\ \mathrm{MeV/c^2}} \else $\mathrm{MeV/c^2}$ \fi}
\def\ps{\ifmmode {\ \mathrm{ps}} \else ps \fi}
\def\pb{ {\ifmmode \;{{\mbox{\mathrm pb}^{-1}}}  \else
         { pb$^{-1}$ } \fi }}
\def\etal{et~al.}
\newcommand {\downto}
         {\mbox{ \begin{picture}(14,10)
                    \put(0,10){\line(0,-1){5.0}}
                    \put(2,5){\oval(4,4)[bl]}
                    \put(1,0){\makebox(0,0)[bl]{$\rightarrow$}}
                 \end{picture} }}
\newcommand{\definmath}[2] {\def#1{\ifmmode#2\else$#2$\fi}}

\def\qr{\ifmmode {{Q_{F}}} \else {${Q_{F}}$}\fi}
\def\ql{\ifmmode {{Q_{B}}} \else {${Q_{B}}$}\fi}
\def\Nl{\ifmmode {{N_{gleich}}} \else {${N_{gleich}}$}\fi}
\def\Nu{\ifmmode {{N_{ungleich}}} \else {${N_{ungleich}}$}\fi}
\def\bmix{\ifmmode {{\mathrm B^{0}}{\mathrm \bar B^{0}}} \else {${\mathrm B^{0}}{\mathrm \bar B^{0}}$}\fi}
\def\bsmix{\ifmmode {{\mathrm B_{s}^{0}}{\mathrm \bar B_{s}^{0}}} \else 
{${\mathrm B_{s}^{0}}{\mathrm \bar B_{s}^{0}}$}\fi}

\definmath{\FB} {\mathrm FB}

\definmath{\X} {{\rm X}}
\definmath{\PWpm} {\mathrm{W}^{\pm}}      % W+-
\definmath{\Pgtp} {\tau^{+}}        % tau+
\definmath{\Pgtm} {\tau^{-}}        % tau-
\definmath{\Pgtpm}   {\tau^{\pm}}         % tau+-
\definmath{\Pgn}  {\nu}          % neutrino
\definmath{\Pagn} {\overline{\nu}}     % anti-neutrino
\definmath{\Pq}      {\mathrm{q}}
\definmath{\PQ}      {\mathrm{Q}}
\definmath{\Paq}  {\overline{\mathrm{q}}}
\definmath{\Pf}      {\mathrm{f}}
\definmath{\Paf}  {\overline{\mathrm{f}}}
\definmath{\Pu}      {\mathrm{u}}
\definmath{\Pau}  {\overline{\mathrm{u}}}
\definmath{\Pd}      {\mathrm{d}}
\definmath{\Pad}  {\overline{\mathrm{d}}}
\definmath{\Ps}      {\mathrm{s}}
\definmath{\Pas}  {\overline{\mathrm{s}}}
\definmath{\Pc}      {\mathrm{c}}
\definmath{\Pac}  {\overline{\mathrm{c}}}
\definmath{\Pb}      {\mathrm{b}}
\definmath{\Pab}  {\overline{\mathrm{b}}}
\definmath{\Pt}      {\mathrm{t}}
\definmath{\Pat}  {\overline{\mathrm{t}}}
\definmath{\Pap}  {\overline{\mathrm{p}}}
\definmath{\Pan}  {\overline{\mathrm{n}}}
\definmath{\PaD}  {\overline{\mathrm{D}}}
\definmath{\PaDz} {\overline{\mathrm{D}}^{0}}
\definmath{\PaB}  {\overline{\mathrm{B}}}
\definmath{\PaBz} {\overline{\mathrm{B}}^{0}}
\definmath{\PaBsz} {\overline{\mathrm{B}}_{\s}^{0}}
\definmath{\PsDpm}   {\mathrm{D}^{\pm}_{\mathrm{s}}}  % Ds+-
\definmath{\PcgLpm}  {\Lambda^{\pm}_{\mathrm{c}}}  % Lambda_c+-
\definmath{\PcgL}  {\Lambda_{\mathrm{c}}}  % Lambda_c
\definmath{\PsBz}  {\overline{\mathrm{B}^0_{\mathrm{s}}}} % Bs0
\definmath{\PbgL}  {\Lambda_{\mathrm{c}}}  % Lambda_c
\definmath{\PKm} {{\rm K^-}}
\definmath{\PKp} {{\rm K^+}}
\definmath{\Pphi} {\phi}            % phi
\definmath{\PD} {\mathrm{D}}     % D
\definmath{\PDp}{{\rm D}^+}
\definmath{\PDm}{{\rm D}^-}
\definmath{\PDs}{{\rm D_s}}    
\definmath{\PDsp}{{\rm D_s^+}}
\definmath{\PDsm}{{\rm D_s^-}}    
\definmath{\PDsst} {{\rm D_s^*}}     % Ds*
\definmath{\PDsstpm} {{\rm D_s^{*\pm}}}     % Ds*+-
\definmath{\PDz} {{\rm D}^{0}}
\definmath{\PaDz} {\overline{\rm D}^{0}}
\definmath{\PDst} {{\rm D}^{*}}     % D*
\definmath{\PDstm} {{\rm D}^{*-}}     % D*-
\definmath{\PDstp} {{\rm D}^{*+}}     % D*+
\definmath{\PDstpm} {{\rm D}^{*\pm}}     % D*+-
\definmath{\PDstz} {{\rm D}^{*0}}     % D*0
\definmath{\PaDstz} {\overline{\rm D}^{*0}}     % D*0
\definmath{\PDzp} {{\rm D}^{0/+}}     % D*
\definmath{\PDstzp} {{\rm D}^{*0/+}}     % D*0/+
\definmath{\PDstzbHuge} {\raisebox{1.7ex}{\tiny( \ \ \ )} \!\!\!\! {\bar{\rm D}}^{*0}}
\definmath{\PDstzb} {\raisebox{1.7ex}{{\tiny(}  {\tiny)}} \!\!\!\!\! {\bar{\rm D}}^{*0}}
\definmath{\PDstst}     {{\rm D}^{**}}       % D**
\definmath{\PDststz}    {{\rm D}^{**0}}      % D**0
\definmath{\PaDststz}    {\overline{\rm D}^{**0}}      % barD**0
\definmath{\PDststp}    {{\rm D}^{**+}}      % D**+
\definmath{\PDststm}    {{\rm D}^{**-}}      % D**-
\definmath{\PDststA}    {{\rm D}_0^{*}}
\definmath{\PDststB}    {{\rm D}_1'}
\definmath{\PDststC}    {{\rm D}_1}
\definmath{\PDststD}    {{\rm D}_2^{*}}
\definmath{\PDststAz}   {{\rm D}_0^{*0}}
\definmath{\PDststBz}   {{\rm D}_1'^0}
\definmath{\PDststCz}   {{\rm D}_1^0}
\definmath{\PDststDz}   {{\rm D}_2^{*0}}
\definmath{\PaDststAz}   {\overline{\rm D}_0^{*0}}
\definmath{\PaDststBz}   {\overline{\rm D}_1'^0}
\definmath{\PaDststCz}   {\overline{\rm D}_1^0}
\definmath{\PaDststDz}   {\overline{\rm D}_2^{*0}}
\definmath{\PDststAp}   {{\rm D}_0^{*+}}
\definmath{\PDststBp}   {{\rm D}_1'^+}
\definmath{\PDststCp}   {{\rm D}_1^+}
\definmath{\PDststDp}   {{\rm D}_2^{*+}}
\definmath{\PDststAm}   {{\rm D}_0^{*-}}
\definmath{\PDststBm}   {{\rm D}_1'^-}
\definmath{\PDststCm}   {{\rm D}_1^-}
\definmath{\PDststDm}   {{\rm D}_2^{*-}}
\definmath{\PDsstst}    {{\rm D_s^{**}}}       % Ds**
\definmath{\PDsststp}   {{\rm D_s^{**+}}}       % Ds**+
\definmath{\PDsststm}   {{\rm D_s^{**-}}}       % Ds**-
\definmath{\PDsststA}  {{\rm D_{s0}^{*}}}
\definmath{\PDsststB}  {{\rm D_{s1}'^}}
\definmath{\PDsststC}  {{\rm D_{s1}^}}
\definmath{\PDsststD}  {{\rm D_{s2}^{*}}}
\definmath{\PDsststAp}  {{\rm D_{s0}^{*+}}}
\definmath{\PDsststBp}  {{\rm D_{s1}'^+}}
\definmath{\PDsststCp}  {{\rm D_{s1}^+}}
\definmath{\PDsststDp}  {{\rm D_{s2}^{*+}}}
\definmath{\PDsststAm}  {{\rm D_{s0}^{*-}}}
\definmath{\PDsststBm}  {{\rm D_{s1}'^-}}
\definmath{\PDsststCm}  {{\rm D_{s1}^-}}
\definmath{\PDsststDm}  {{\rm D_{s2}^{*-}}}
\definmath{\PaDstz} {\overline{{\rm D}^{*0}}}     % D*0bar
\definmath{\PDsstm} {{\rm D_s^{*-}}}     % D_s*-
\definmath{\PDsstp} {{\rm D_s^{*+}}}     % D_s*+

\definmath{\PBz} {\mathrm{B}^0}     % B0
\definmath{\PBp} {\mathrm{B}^+}     % B+
\definmath{\PBm} {\mathrm{B}^-}     % B-
\definmath{\PBsz} {\mathrm{B}_s^0}     % B_s0
\definmath{\PBst} {{\rm B}^*} % B*
\definmath{\PBstst} {{\rm B}^{**}} % B**

\definmath{\Ppiz} {\pi^0{}}
\definmath{\Ppip} {\pi^+{}}
\definmath{\Ppim} {\pi^-{}}
\definmath{\Ppipm} {\pi^\pm}
\definmath{\Pepm} {\mathrm{e}^\pm}
\definmath{\PZz} {{\rm Z^0}}

\definmath{\PgLz} {\Lambda^{0}}        % Lambda0
\def\D0{\ifmmode {{\mathrm D^0}} \else {${\mathrm D^0}$}\fi}
\def\dEdx { \ifmmode {\rm dE/dx} \else {dE/dx} \fi}
\newcommand{\NIM}{Nucl.~Instr.~and~Meth.}
\newcommand{\PR}{Phys.~Rev.}
\newcommand{\PL}{Phys.~Lett.}
\newcommand{\PRL}{Phys.~Rev.~Lett.}
\newcommand{\ZPhys}{Z.~Phys.}

\definmath{\cm} {\mathrm{cm}}
\definmath{\mm} {\mathrm{mm}}
\definmath{\m} {\mathrm{m}}
\definmath{\mrad} {\mathrm{mrad}}

\newcommand{\MC} {Monte Carlo}
\newcommand{\ie} {i.\,e.}
\definmath{\opa} {<\hspace{-7pt})\,}
\definmath{\einhalb} {{}^{1\!\hspace{-1pt}}/_{\!2}}
\definmath{\RR} {{\textstyle\raisebox{5.25pt}{-}\hspace{-4pt}\raisebox{-2.5pt}{-}\hspace{-4pt}\raisebox{2pt}{$\scriptstyle |$}\hspace{-2pt}{\rm R}}}
\definmath{\RRsmall} {{\textstyle\raisebox{4.8pt}{-}\hspace{-4pt}\raisebox{-2.35pt}{-}\hspace{-3.7pt}\raisebox{1.9pt}{$\scriptstyle |$}\hspace{-1.9pt}{\rm R}}}

\definmath{\thetastarpiD} {\theta^*_{\pi^+\!,\,\rm D}}
\definmath{\thetastarpiDz} {\theta^*}          % _{\pi^+\!,\,\rm D^0}}
\definmath{\thetastarpiDsp} {\theta^*_{\pi^+\!,\,\rm D_s^+}}

\definmath{\stat} {\textstyle{\it{(\!stat\hspace{-1pt}.\hspace{-1pt})}}}
\definmath{\syst} {\textstyle{\it{(\!syst\hspace{-1pt}.\hspace{-1pt})}}}
\definmath{\ext} {\textstyle{\it{(\!ext\hspace{-1pt}.\hspace{-1pt})}}}
\definmath{\extrap} {\textstyle{\it{(\!extrap\hspace{-1pt}.\hspace{-1pt})}}}

\definmath{\VVP} {P_V}
\definmath{\VVPF} {P_V^{\rm eff}}
\definmath{\VVPP} {P_V^{\rm prim}}
\definmath{\VVPPs} {P_V^{\rm prim,\,s}} 
\definmath{\Nc} {N^{\rm c}}
\definmath{\Pr} {Pr}

\definmath{\BrDstp} {Br(\PDstp\!\to\!\PDz\Ppip)}
\definmath{\BrDz} {Br(\PDz\!\to\!\PKm\Ppip)}
\definmath{\BrDp} {Br(\PDp\!\to\!\PKm\Ppip\Ppip)}
\definmath{\BrDsp} {Br(\PDsp\!\to\!\Pphi\Ppip)}
\definmath{\BrPhi} {Br(\Pphi\!\to\!\PKp\PKm)}
\definmath{\BrDstzgammapiz} 
                  {Br(\PDstz\!\to\!\PDz\gamma)/Br(\PDstz\!\to\!\PDz\Ppiz)}
\definmath{\BrDstpshort} {Br_{\PDstp}}
\definmath{\BrDzshort} {Br_{\PDz}}
\definmath{\BrDpshort} {Br_{\PDp}}
\definmath{\BrDspshort} {Br_{\PDsp}}
\definmath{\BrPhishort} {Br_{\Pphi}}
\definmath{\fPDst} {f(\c\!\to\!\PDst)}
\definmath{\fPDstshort} {f_{\PDst}}
\definmath{\fPDstz} {f(\c\!\to\!\PDstz)}
\definmath{\fPDstzshort} {f_{\PDstz}}
\definmath{\fPDstp} {f(\c\!\to\!\PDstp)}
\definmath{\fPDstpshort} {f_{\PDstp}}
\definmath{\fPDsstp} {f(\c\!\to\!\PDsstp)}
\definmath{\fPDsstpshort} {f_{\PDsstp}}
\definmath{\fPDz} {f(\c\!\to\!\PDz)}
\definmath{\fPDzshort} {f_{\PDz}}
\definmath{\fPDp} {f(\c\!\to\!\PDp)}
\definmath{\fPDpshort} {f_{\PDp}}
\definmath{\fPDsp} {f(\c\!\to\!\PDsp)}
\definmath{\fPDspshort} {f_{\PDsp}}
\definmath{\fPDzp} {f(\c\!\to\!\PDzp)}
\definmath{\fPDzpshort} {f_{\PDzp}}
\definmath{\fPDstzp} {f(\c\!\to\!\PDstzp)}
\definmath{\fPDstzpshort} {f_{\PDstzp}}
\definmath{\fPDststA} {f(\c\!\to\!\PDststA)}
\definmath{\fPDststz} {f(\c\!\to\!\PDststz)}
\definmath{\fPDststp} {f(\c\!\to\!\PDststp)}
\definmath{\fPDsststp} {f(\c\!\to\!\PDsststp)}
\definmath{\fPDststAshort} {f_{\PDststA}}
\definmath{\fPDststB} {f(\c\!\to\!\PDststB)}
\definmath{\fPDststBshort} {f_{\PDststB}}
\definmath{\fPDststC} {f(\c\!\to\!\PDststC)}
\definmath{\fPDststCshort} {f_{\PDststC}}
\definmath{\fPDststD} {f(\c\!\to\!\PDststD)}
\definmath{\fPDststDshort} {f_{\PDststD}}
\definmath{\fPDststAz} {f(\c\!\to\!\PDststAz)}
\definmath{\fPDststAzshort} {f_{\PDststAz}}
\definmath{\fPDststBz} {f(\c\!\to\!\PDststBz)}
\definmath{\fPDststBzshort} {f_{\PDststBz}}
\definmath{\fPDststCz} {f(\c\!\to\!\PDststCz)}
\definmath{\fPDststCzshort} {f_{\PDststCz}}
\definmath{\fPDststDz} {f(\c\!\to\!\PDststDz)}
\definmath{\fPDststDzshort} {f_{\PDststDz}}
\definmath{\fPDsststAp} {f(\c\!\to\!\PDsststAp)}
\definmath{\fPDsststApshort} {f_{\PDsststAp}}
\definmath{\fPDsststBp} {f(\c\!\to\!\PDsststBp)}
\definmath{\fPDsststBpshort} {f_{\PDsststBp}}
\definmath{\fPDsststCp} {f(\c\!\to\!\PDsststCp)}
\definmath{\fPDsststCpshort} {f_{\PDsststCp}}
\definmath{\fPDsststDp} {f(\c\!\to\!\PDsststDp)}
\definmath{\fPDsststDpshort} {f_{\PDsststDp}}
\definmath{\fbPDst} {f(\b\!\to\!\PDst)}
\definmath{\fbPDstz} {f(\b\!\to\!\PDstz)}
\definmath{\fbPDstp} {f(\b\!\to\!\PDstp)}
\definmath{\fbPDsstp} {f(\b\!\to\!\PDsstp)}
\definmath{\fbPDz} {f(\b\!\to\!\PDz)}
\definmath{\fbPDp} {f(\b\!\to\!\PDp)}
\definmath{\fbPDsp} {f(\b\!\to\!\PDsp)}
\definmath{\fbPDststAz} {f(\b\!\to\!\PDststAz)}
\definmath{\fbPDststBz} {f(\b\!\to\!\PDststBz)}
\definmath{\fbPDststCz} {f(\b\!\to\!\PDststCz)}
\definmath{\fbPDststDz} {f(\b\!\to\!\PDststDz)}
\definmath{\fbPDsststAp} {f(\b\!\to\!\PDsststAp)}
\definmath{\fbPDsststBp} {f(\b\!\to\!\PDsststBp)}
\definmath{\fbPDsststCp} {f(\b\!\to\!\PDsststCp)}
\definmath{\fbPDsststDp} {f(\b\!\to\!\PDsststDp)}
\definmath{\subxDz} {\,\rule[-0.8ex]{0.04pt}{2.6ex}_{\,x_{\PDz}>0.3}}
\definmath{\subxDsp} {\,\rule[-0.8ex]{0.04pt}{2.6ex}_{\,x_{\PDsp}>0.35}}
\definmath{\fxPDstz} {f(\c\!\to\!\PDstz)^{x_{\PDz}>0.3}}
\definmath{\fxPDstp} {f(\c\!\to\!\PDstp)^{x_{\PDz}>0.3}}
\definmath{\fxPDsstp} {f(\c\!\to\!\PDsstp)^{x_{\PDsp}>0.35}}
\definmath{\fxPDstzp} {f(\c\!\to\!\PDstzp)^{x_{\PDz}>0.3}}
\definmath{\fxbPDstz} {f(\b\!\to\!\PDstz)^{x_{\PDz}>0.3}}
\definmath{\fxbPDstp} {f(\b\!\to\!\PDstp)^{x_{\PDz}>0.3}}
\definmath{\fxbPDsstp} {this macro does not work!f(\b\!\to\!\PDsstp)^{x_{\PDsp}>0.35}}

\definmath{\etadat} {\eta_{\rm dat}}
\definmath{\etamc}  {\eta_{\rm MC}}
\definmath{\Fc}     {{\cal F}_\c}
\definmath{\FxcPDstz}     {{\cal F}_\c(\PDstz)^{x_{\PDz}>0.3}}
\definmath{\FxcPDstp}     {{\cal F}_\c(\PDstp)^{x_{\PDz}>0.3}}
\definmath{\FxcPDsstp}    {{\cal F}_\c(\PDsstp)^{x_{\PDsp}>0.3}}
\definmath{\FxcPDstzp}     {{\cal F}_\c(\PDstzp)^{x_{\PDz}>0.3}}
\definmath{\Fb}     {{\cal F}_\b}
\definmath{\FxbPDstz}     {{\cal F}_\b(\PDstz)^{x_{\PDz}>0.3}}
\definmath{\FxbPDstp}     {{\cal F}_\b(\PDstp)^{x_{\PDz}>0.3}}
\definmath{\FxbPDsstp}    {{\cal F}_\b(\PDsstp)^{x_{\PDsp}>0.3}}

\definmath{\plhmpq} {p_{\rm frag}^l}

\begin{document}
% define rotations for axis labels:
  \def\rotninety{\special{ps:currentpoint currentpoint translate 90 rotate neg exch
                             neg exch translate}}
  \def\rotmnine{ \special{ps:currentpoint currentpoint translate -90 rotate neg exch
                             neg exch translate}
               }        
  \def\rottwoseven{
                   \special{ps:currentpoint currentpoint translate 270 rotate neg exch
                               neg exch translate}
                   }         
% (use as follows:                                                         
% \special{ps:gsave}
% \put(-2.5,11.6){\rotninety\rotninety\rotninety Number of entries / 0.5 MeV}
% \special{ps:grestore}                  
% )

\abovedisplayskip 4pt %minus 2pt plus 2pt
\belowdisplayskip 4pt %minus 2pt plus 2pt

\setlength{\parskip}{0.7ex plus 0.2ex minus 0.2ex}
\setlength{\itemsep}{0.0ex plus 0.0ex minus 0.0ex}
\setlength{\topsep}{0.0ex plus 0.0ex minus 0.0ex}
\setlength{\partopsep}{0.0ex plus 0.0ex minus 0.0ex}
\setlength{\parsep}{0.0ex plus 0.0ex minus 0.0ex}
\renewcommand{\textfraction}{0.05}
\renewcommand{\topfraction}{0.95}
\renewcommand{\bottomfraction}{0.95}

\definmath{\VNtkmhwomillion}{
4.32
}
\definmath{\VNmhmcwomillion}{
6.5
}

\definmath{\VrawfitresDstzgsg}{
148 \pm 25
}
\definmath{\VrawfitresDstzgsp}{
139 \pm 27
}
\definmath{\VrawfitresDstzpsp}{
110 \pm 13
}
\definmath{\VrawfitresDsstpgsg}{
132 \pm 40
}

\definmath{\VchisqDstz}{
61.6
}
\definmath{\VdofDstz}{
69
}
\definmath{\VdatachisqDspgamma}{
37.1
}
\definmath{\VdofDspgamma}{
43
}
\definmath{\VmcchisqDspgamma}{
23.9
}
\definmath{\VdatachisqDsppiz}{
1.62
}
\definmath{\VdataKolmogDsppiz}{
62\%
}
\definmath{\VmcchisqDsppiz}{
0.64
}
\definmath{\VmcKolmogDsppiz}{
96\%
}

\definmath{\VeffsigDstzcgsg}{
( 2.88 \pm 0.16 ) \%
}
\definmath{\VeffsigDstzbgsg}{
( 2.15 \pm 0.19 ) \%
}
\definmath{\VeffsigDstzcgsp}{
( 1.80 \pm 0.10 ) \%
}
\definmath{\VeffsigDstzbgsp}{
( 1.01 \pm 0.10 ) \%
}
\definmath{\VeffsigDstzcpsp}{
( 1.18 \pm 0.08 ) \%
}
\definmath{\VeffsigDstzbpsp}{
( 1.05 \pm 0.10 ) \%
}
\definmath{\VeffsigDsstpcgsg}{
( 6.05 \pm 0.55 ) \%
}
\definmath{\VeffsigDsstpbgsg}{
( 6.47 \pm 0.64 ) \%
}

\definmath{\VetaReffsigbkgDzgamma}{
0.20 \pm 0.03
}
\definmath{\VetaReffsigbkgDzpiz}{
0.16 \pm 0.01
}
\definmath{\VetaReffsigbkgDspgamma}{
0.09
}

\definmath{\VrestDstz}{
24.9 \%
}
\definmath{\VrestDsstp}{
27.4\%
}
\definmath{\VresybsubDstz}{
12.2\%
}
\definmath{\VresyDstz}{
20.9 \%
}
\definmath{\VresyxDsstp}{
23.8\%
}
\definmath{\VresyDsstp}{
30.8\%
}

\definmath{\VnbarxPDstzBr}{
(3.92 \pm 0.56 \,\stat \pm 0.66 \,\syst ) \times 10^{-3}
}
\definmath{\VnbarxPDsstpBr}{
(7.1 \pm 1.9 \pm 1.7 ) \times 10^{-4}
}
\definmath{\VnbarxPDsstpBrsummary}{
(7.1 \pm 1.9 \,\stat \pm 1.7 \,\syst ) \times 10^{-4}
}
\definmath{\VnbarPDsstpBr}{
(1.69 \pm 0.46 \pm 0.40 \pm 0.33 ) \times 10^{-3}
}
\definmath{\VnbarPDsstpBrsummary}{
(1.69 \pm 0.46 \,\stat \pm 0.52 \,\syst ) \times 10^{-3}
}

\definmath{\VfPDstzxBr}{
(1.13 \pm 0.28 \,\stat \pm 0.23 \,\syst ) \!\times\! 10^{-3}
}
\definmath{\VfPDsstpxBr}{
(1.27 \pm 0.87 \,\stat \pm 0.56 \,\syst ) \!\times\! 10^{-4}
}

\definmath{\VfPDstzBrstat}{
(1.44 \pm 0.36 ) \!\times\! 10^{-3}
}

\definmath{\VfPDstzBr}{
(1.44 \pm 0.36 \,\stat \pm 0.30 \,\syst ) \!\times\! 10^{-3}
}
\definmath{\VfPDsstpBr}{
(1.70 \pm 1.16 \,\stat \pm 0.76 \,\syst ) \!\times\! 10^{-4}
}

\definmath{\VfPDstzBrfortable}{
(1.44 \pm 0.36 \pm 0.30 ) \!\times\! 10^{-3}
}
\definmath{\VfPDsstpBrfortable}{
(1.70 \pm 1.16 \pm 0.76 ) \!\times\! 10^{-4}
}

\definmath{\VfPDstz}{
0.218 \pm 0.054 \,\stat \pm 0.045 \,\syst \pm 0.007 \,\ext 
}
\definmath{\VfPDsstp}{
0.059 \pm 0.040 \,\stat \pm 0.026 \,\syst \pm 0.015 \,\ext 
}

\definmath{\VfPDstzabstract}{
0.218 \pm 0.054 \pm 0.045 \pm 0.007 
}
\definmath{\VfPDsstpabstract}{
0.059 \pm 0.040 \pm 0.026 \pm 0.015 
}

\definmath{\VBrgammapi0}{
0.39 \pm 0.17 \,\stat
}

\definmath{\VVVPF}{
0.57 \pm 0.05
}

\definmath{\VVVPO}{
0.55 \pm 0.10
}

\definmath{\VfPDstpBr}{
(1.041 \pm 0.020 \pm 0.040) \times 10^{-3}
}
\definmath{\VbetaxPDstp}{
( 0.980 \pm 0.068 ) \times 10^{-3}
}
\definmath{\VfbPDstpBr}{
(1.334 \pm 0.049 \pm 0.078) \times 10^{-3}
}

\definmath{\VbetaxPDsstp}{
(1.82 \pm 0.49) \times 10^{-4}
}

\definmath{\VspinalPA}{
0.23 \pm 0.02
}
\definmath{\VspinalPB}{
0.34 \pm 0.03
}
\definmath{\VspinalPC}{
0.43 \pm 0.05
}

\definmath{\VT}{
\left( 174^{+62}_{-36} \right) \MeV
}

\definmath{\VfPDstzPDstp}{
0.94 \pm 0.31
}
\definmath{\VRiso}{
1.14 \pm 0.23
}
\definmath{\VRjso}{
1.19 \pm 0.36
}

\definmath{\VfPDstBr}{
(1.045 \pm 0.044 ) \times 10^{-3}
}

\definmath{\Vgammas}{
0.23 \pm 0.19
}

\definmath{\VVVPPnospincounting}{
0.55 \pm 0.08
}

\definmath{\VnsigmaVVPP}{
2.7
} 

\definmath{\VA}{
(2.29 \pm 0.34 ) \times 10^{-2}
}
\definmath{\VB}{
(3.82 \pm 0.89 ) \times 10^{-2}
}
\definmath{\VABcorrel}{
77 \%
}

\definmath{\VVVPPs}{
0.58 \pm 0.50
}

\definmath{\VVVPPlowest}{
0.48 \pm 0.09
}
\definmath{\VVVPPhighest}{
0.65 \pm 0.11
}

\renewcommand{\Huge}{\huge}
\parskip12pt plus 1pt minus 1pt
\topsep0pt plus 1pt
\begin{titlepage}
\begin{center}{\large   EUROPEAN LABORATORY FOR PARTICLE PHYSICS
}\end{center}\bigskip
\begin{flushright}
CERN-EP/98-006\\
12 January 1998\\
\end{flushright}

\bigskip

\begin{center}{\LARGE\bf   
\boldmath
Determination of the Production Rate\\ 
of $\PDstz$ Mesons and of the Ratio $V\!$/($V$+$P$)\\
 in $\PZz\!\to\!\cc$ Decays
\unboldmath
}\end{center}\bigskip\bigskip
\begin{center}{\Large The OPAL Collaboration
}\end{center}\bigskip\bigskip
%\centerline{\Large Authors: Frank Fiedler, Johannes Steuerer}
%\centerline{\Large Editorial Board: Ties Behnke, Christoph Burgard,} 
%\centerline{\Large Jules Gascon, Dale Koetke, and Alex Martin}
\bigskip\begin{center}{\large  Abstract}\end{center}
\noindent
In $\epem$ collisions at centre-of-mass energies around $91$ GeV,
$\PDstz$ mesons 
have been reconstructed 
using data collected with the OPAL detector at LEP.
The hadronisation fraction has been 
measured to be
\[
\begin{array}{r@{\ =\ }l}
\protect{\fPDstz} &
\protect{\VfPDstzabstract}
\,,
\end{array}
\]
where the errors correspond to the statistical and systematic errors
specific to this analysis, and to
systematic uncertainties from externally measured branching fractions,
respectively.
Together with previous OPAL measurements of the hadronisation fractions of 
other charmed mesons, this value is used 
to investigate
the relative production of observed vector and
pseudoscalar charmed mesons in $\PZz\!\to\!\cc$ decays.
The production ratio is determined to be
\[
\VVPF = V/(V\!+\!P) =
\protect{\VVVPF}
\ .
\]
The relative primary production of vector and pseudoscalar mesons,
$\VVPP$, is studied in the context 
of the production and decay of orbitally excited charmed resonances.
The first measurement of the inclusive $\PDsstp$ production rate in hadronic
$\PZz$ decays is presented.

\vfill

\begin{center}{\large
(Submitted to Eur.~Phys.~J.~C)}
\end{center}
\end{titlepage} 
\begin{center}{\Large        The OPAL Collaboration
}\end{center}\bigskip
\begin{center}{
%begin authorlist
K.\thinspace Ackerstaff$^{  8}$,
G.\thinspace Alexander$^{ 23}$,
J.\thinspace Allison$^{ 16}$,
N.\thinspace Altekamp$^{  5}$,
K.J.\thinspace Anderson$^{  9}$,
S.\thinspace Anderson$^{ 12}$,
S.\thinspace Arcelli$^{  2}$,
S.\thinspace Asai$^{ 24}$,
S.F.\thinspace Ashby$^{  1}$,
D.\thinspace Axen$^{ 29}$
% ...
%end authorlist
}\end{center}\bigskip
\bigskip
%begin institutes
$^{  1}$School of Physics and Astronomy, University of Birmingham,
Birmingham B15 2TT, UK
\newline
$^{  2}$Dipartimento di Fisica dell' Universit\`a di Bologna and INFN,
I-40126 Bologna, Italy
\newline
% ...
$^{  5}$Cavendish Laboratory, Cambridge CB3 0HE, UK
\newline
% ...
$^{  8}$CERN, European Organisation for Particle Physics,
CH-1211 Geneva 23, Switzerland
\newline
$^{  9}$Enrico Fermi Institute and Department of Physics,
University of Chicago, Chicago IL 60637, USA
\newline
% ...
$^{ 12}$Indiana University, Department of Physics,
Swain Hall West 117, Bloomington IN 47405, USA
\newline
% ...
$^{ 16}$Department of Physics, Schuster Laboratory, The University,
Manchester M13 9PL, UK
\newline
% ...
$^{ 23}$Department of Physics and Astronomy, Tel Aviv University,
Tel Aviv 69978, Israel
\newline
$^{ 24}$International Centre for Elementary Particle Physics and
Department of Physics, University of Tokyo, Tokyo 113, and
Kobe University, Kobe 657, Japan
\newline
% ...
$^{ 29}$University of British Columbia, Department of Physics,
Vancouver BC V6T 1Z1, Canada
\newline
% ...
\newline
%end notes
%

\newpage

\section{Introduction}
\label{introduction.sec}

A meson with no orbital angular momentum can be a vector ($V$) state with
spin~1 or a pseudoscalar ($P$) state with spin~0.
Their relative production $\VVP$=$V/(V$+$P)$ 
is sensitive to non-perturbative effects in the hadronisation process 
and cannot be calculated exactly.
However, several models~\cite{bib-fragmodels} have been proposed which 
predict this ratio.
A simple spin counting picture, where the abundance of a particular 
state is proportional to its number of spin degrees of freedom\footnote{
``Spin counting'' in this paper refers to a model where the relative 
primary production of mesons with the same quark content and the same 
orbital angular momentum is given according to the corresponding numbers of 
spin degrees of freedom.
}, predicts a value of $\VVP$=$0.75$ when only vector and pseudoscalar 
meson production are considered.
More
sophisticated models take into account the mass difference between
vector and pseudoscalar mesons, as well as the masses of
the constituent quarks. 
In general, they predict 
$\VVP$ to be less than $0.75$.

The situation is complicated by the presence of mesons 
with non-zero orbital angular momentum. 
The observed $\VVP$ ratio 
thus includes those ground state mesons which have 
been produced in decays of the excited states. 
Experimentally,
this effective ratio $\VVPF$ is easier to measure. 
If the production rates of the excited states are known,
the primary ratio $\VVPP$, corrected for any effects 
from excited states, can be calculated. 
The LEP accelerator, where numerous $\PZz\!\to\!\qq$ decays have been observed
between 1989 and 1995, provides a facility to study the $\VVP$ ratio.
Values around $\VVPF$$\approx$$0.75$ and
$\VVPF$$\approx$$0.55$ have been measured
for B meson production~\cite{bib-VVPBstar} and
in the charm sector~\cite{bib-christoph,bib-VVPvalues},
respectively. In both cases, a model dependent evaluation suggests
the values of $\VVPP$ and $\VVPF$ to be similar~\cite{bib-rostock}.
For light mesons, $\VVPF$ has been estimated to be
between $0.4$ and $0.5$~\cite{bib-pei}.
This paper focuses on a study of both $\VVPF$ and $\VVPP$ for charmed
meson production.

In the charm system, measurements of $\PDz$, $\PDp$, and $\PDstp$ production in 
$\PZz\!\to\!\cc$ decays have been used in the previous determinations of 
$\VVPF$~\cite{bib-christoph,bib-VVPvalues}.
The $\PDstz$ meson has so far not been observed in $\PZz$ decays,
since it only decays via the emission of a photon or a 
$\Ppiz$ meson, which are difficult to reconstruct experimentally.
Therefore, the determinations of $\VVP$ values
have so far relied on the assumption of isospin 
invariance, which suggests equal $\PDstz$ and $\PDstp$ production rates.

In this paper, a first measurement of the hadronisation fraction 
$\fPDstz$ in $\PZz\!\to\!\cc$ decays is presented.
The analysis is based on more than 4 million hadronic $\PZz$ decays recorded
with the OPAL detector at the LEP accelerator in the years 1990 to 1995.
The $\PDstz$ mesons are reconstructed in 
the decay channels $\PDstz\!\to\!\PDz\gamma$ and $\PDstz\!\to\!\PDz\Ppiz$.
The same techniques are applied in the $\PDsp\gamma$ final state for the 
reconstruction of $\PDsstp$ mesons, which at LEP have only been observed
in leptonic $\PDsp$
decays~\cite{bib-PDsstp_at_L3}.
The $\fPDstz$ measurement is used to test the assumption of 
isospin invariance.
Together with previously published OPAL measurements of 
the other charmed non-strange pseudoscalar and vector
mesons~\cite{bib-stableDcounting,bib-Tiesdstarplus},
a value of $\VVPF$ is derived.
A recent OPAL measurement of the production of excited charmed 
mesons~\cite{bib-Ddoublestar} is used to investigate the ratio $\VVPP$.
A model independent formula
for $\VVPP$ is derived, and the validity of a simple spin counting model
for the fragmentation process is tested.

The paper is organized as follows: Section~\ref{opal.sec} contains a 
brief description of the OPAL detector and the event selection. 
Section~\ref{reconstruction.sec}
describes the reconstruction of $\PDstz$ decays,
the selection of signal candidates, and the background determination.  
In section~\ref{systunc.sec}, the measurement of the 
hadronisation fraction $\fPDstz$ and its
systematic uncertainties are discussed, while section~\ref{dsstp.sec} describes
the measurement of $\PDsstp$ production.
Section~\ref{vvp.sec} contains
a determination of both $\VVPF$ and $\VVPP$ for charmed meson production in
$\PZz\!\to\!\cc$ decays and an interpretation of the results.

\section{The OPAL Detector and Event Selection}
\label{opal.sec}

A complete description of the OPAL detector is given 
elsewhere~\cite{bib-elsewhere}. Here, only the components of importance for 
this analysis are reviewed.
Tracking of charged particles is performed by 
a silicon microvertex detector, 
a vertex detector, a jet chamber, and a set of drift chambers 
that measure the coordinate of tracks along the direction of 
the beam line\footnote{
The OPAL coordinate system is defined with positive $z$ along the electron 
beam direction and the $x$ axis horizontal; $\theta$ and $\phi$ are the polar 
and azimuthal angles, respectively.
} ($z$-chambers),
positioned inside a solenoid that provides a 
uniform magnetic field of 0.435 T parallel to the beam direction. 

The barrel electromagnetic calorimeter, which
covers the polar angle range of $|\cos\theta|<0.82$, is mounted outside 
the magnet coil.
It consists of a cylindrical array of 9,440 lead glass blocks 
of 24.6~radiation lengths thickness pointing approximately 
to the interaction region. 
The overall energy resolution is improved by
correcting for the energy lost in showers initiated in the
material in front of the calorimeter.
Such showers are detected by thin gas detectors (presampler detectors)
situated in front of the lead glass blocks, and by
time-of-flight 
scintillators located between the presampler and
the magnet coil in the polar angle range $|\cos\theta|<0.72$.
The regions $0.82<|\cos\theta|<0.98$ are covered by the endcap 
electromagnetic calorimeters with lead glass blocks
oriented parallel to the beam direction.
The magnet return yoke is instrumented as a hadron calorimeter. 
Four layers of muon chambers are mounted outside the hadron calorimeter.

The criteria for selecting hadronic $\PZz$ decays are based on 
reconstructed tracks in the central detector and on the energy distribution 
in the calorimeter~\cite{bib-MHsel}. Charged particle 
tracks need to have at least 20
jet chamber hits, a momentum component in the $xy$ plane
of at least $0.15\,\GeV$, a total momentum of less than
$65\,\GeV$, and a distance of closest approach to the beam axis of less
than 5\,cm.
The hadronic $\PZz$~event selection efficiency of these requirements
is $(98.7\pm0.4)\%$~\cite{bib-Tiesdstarplus}.
Of the events recorded with the OPAL
detector between 1990 and 1995, $\VNtkmhwomillion$ million satisfy the
event selection criteria.
The primary vertex is
reconstructed from the charged tracks in the event and
constrained with the known
average beam position and the $xy$ width of the $\epem$ collision point.
 
Samples of simulated hadronic events
are used for the determination of selection criteria and 
for the calculation of selection efficiencies. 
They have been generated using the 
\mbox{JETSET $7.4$} \MC{ }model~\cite{bib-JETSET} 
with parameters tuned to reproduce the OPAL data~\cite{bib-OPALtune}.
The fragmentation of heavy quarks is parametrised by
the fragmentation function of Peterson \etal~\cite{bib-peterson}.
The simulated events 
are then processed by the detector simulation program~\cite{bib-GOPAL}
and by the same reconstruction algorithm which is also applied to the data.
%        -------------------
\section{Reconstruction and Selection of 
         \boldmath${\rm D}^{*0}$\unboldmath{ }Candidates}
%        -------------------
\label{reconstruction.sec}

In $\PZz$ decays, $\PDstz$ mesons are dominantly
produced in $\PZz\!\to\!\cc$ events and in the decay
of bottom hadrons. 
The $\PDstz$ mesons are reconstructed in the decay modes
$\PDstz\!\to\!\PDz\gamma$ and $\PDstz\!\to\!\PDz\Ppiz$.
The transition photon or $\Ppiz$ meson is 
expected to be dominantly produced in the core of the jet that contains the
$\PDstz$.
Thus, for the reconstruction of $\PDstz\!\to\!\PDz\gamma$ decays,
a considerable background level is expected from 
photons originating in decays of $\Ppiz$ mesons from other sources,
while the reconstruction of 
$\PDstz\!\to\!\PDz\Ppiz$ decays suffers from combinatorial
background in the $\Ppiz\!\to\!\gamma\gamma$ reconstruction.
The reconstruction proceeds by first finding $\PDz$ candidates, which are
then combined with a photon or a $\Ppiz$.
After a loose preselection, a likelihood method is used for the final
selection of candidates.
In this section, 
the selection 
and the
background determination are described. 

%        -------------------
\subsection{Reconstruction 
               of \boldmath${\rm D^0}$\unboldmath{ }Mesons,
               Photons, and \boldmath$\pi^0$\unboldmath{ }Mesons}
%        -------------------
\label{dpresel.subsec}

Both the preselection for the $\PDz$, photon, and
$\Ppiz$ reconstruction and the likelihood for the $\PDstz$ selection
are based on the variables which are explained below.

The decay mode $\PDz\!\to\!\PKm\Ppip$ is used to reconstruct $\PDz$ mesons.
Initially, all possible pairs of oppositely charged tracks are formed,
assigning kaon and pion masses to the tracks.
Variables for the selection of $\PDz$ candidates are:
\begin{list}{$\bullet$}{\itemindent=0pt \topsep=0pt
              \parskip=0pt \parsep=0pt \itemsep=0pt 
              \partopsep=0pt}
\item the invariant mass $m_{\PDz}$ 
and the scaled energy $x_{\PDz}=E_{\PDz}/E_{\rm beam}$ of the $\PDz$ candidate;
\item the cosine of the helicity angle $\thetastarpiDz$ measured between the 
direction of the charged pion in the $\PDz$ rest frame and the 
$\PDz$ direction in the laboratory frame; 
\item the signed probability $W_{\rm X}$ as defined in~\cite{bib-dEdx} 
that a given track 
is compatible with the particle hypothesis ${\rm X}$, based on its 
specific energy loss and its measured momentum;
\item the signed decay length $d_{xy}$ in the $xy$ plane defined as the 
distance between the primary vertex and the secondary vertex formed by 
the $\PDz$ decay products; and
\item the largest longitudinal momentum $\plhmpq$ relative to the $\PDz$ 
flight direction of 
any track which is 
not used in the $\PDz$ reconstruction and whose charge is 
inconsistent with that of an accompanying hadron of the $\PDz$ 
candidate.
This variable helps to separate signal candidates in $\PZz\!\to\!\cc$ events
from background, since in the formation of 
$\PDz$ mesons in $\PZz\!\to\!\cc$ events, 
there is a correlation between the charge of the primary quark and that
of the fragmentation
particle with the highest longitudinal momentum~\cite{bib-Bdoublestar}.
The track with the largest longitudinal momentum whose charge is inconsistent
with being the accompanying hadron is expected to be softer for signal
than for background.
\end{list}

Contributions from $\PDstp$ decays are suppressed by searching
for a track that could have been the pion in a 
$\PDstp\!\to\!\PDz\Ppip$ decay. 
The corresponding $\PDz$ candidate is rejected if a track is found 
with the correct charge and a mass difference within
\mbox{$141\MeV<m_{\rm D^0 \pi^+}$$-$$m_{\rm D^0}<152\MeV$.}

A loose preselection of $\PDz$ candidates is done using the invariant mass
$m_{\PDz}$, the scaled energy $x_{\PDz}$, the helicity angle 
$\cos\thetastarpiDz$, and 
the particle identification probabilities. The exact cuts are listed in
table~\ref{preselcuts.table}.

Photons are reconstructed either as showers in the electromagnetic 
calorimeter, or through conversions $\gamma\!\to\!\epem$. For the latter,
pairs of oppositely charged tracks are identified as 
conversions with the algorithm described in \cite{bib-sigma}.
The $z$ components of the track momenta are determined in a fit to the 
conversion point to improve the momentum resolution.
The total photon momentum is taken 
as the sum of both track momenta.

In the barrel region of the electromagnetic calorimeter,
photons are identified as described in~\cite{bib-sigma} 
by fitting electromagnetic showers to
energy deposits which are not associated with any charged particle track.
The lateral shower profile is fixed to the
\MC{ }expectation, and the normalisation gives the shower energy.
The fitted shower energy is then corrected 
for losses in the material in front of the calorimeter
using information from the presampler and time-of-flight system.
Each of the fitted showers is treated as a photon candidate.
The photon momenta are calculated assuming that the photons
originated at the primary vertex of the event.

Any pair of photons (showers or conversions) with an invariant
mass between $60\MeV$ and $280\MeV$ is considered as a $\Ppiz$ candidate.
The resolution of the $\Ppiz$ energy is improved by a constraint to the 
nominal $\Ppiz$ mass \cite{bib-PDG96} using a kinematic fit~\cite{bib-sigma}.

A number of variables similar to those in reference~\cite{bib-sigma} are
calculated for each photon and $\Ppiz$ candidate:
\begin{list}{$\bullet$}{\itemindent=0pt \topsep=0pt
              \parskip=0pt \parsep=0pt \itemsep=0pt 
              \partopsep=0pt}
\item the energy of the photon or $\Ppiz$ candidate, denoted
$E_\gamma$ or $E_{\Ppiz}$, respectively; 
\item for photon conversions, the signed electron identification probabilities 
$W_{\rm e}$ calculated from the energy loss of the two conversion tracks;
\item for showers in the calorimeter, a variable which parametrises 
how well the fitted shower describes the measured energy distribution
in the 9 calorimeter blocks around the fitted shower 
maximum; and
three variables that describe their effective separation 
from the closest neighbouring shower and the closest charged track
entering the calorimeter.
\end{list}
The following variables are only used in the selection of $\Ppiz$ mesons:
\begin{list}{$\bullet$}{\itemindent=0pt \topsep=0pt
              \parskip=0pt \parsep=0pt \itemsep=0pt 
              \partopsep=0pt}
\item the invariant mass of the $\gamma \gamma$ system and 
the opening angle between the two photons; as well as
\item the number of additional 
photon candidates in a cone around each of the two photon
candidates under consideration. In each case, the opening angle of the cone is
twice the angle between the two photons.
\end{list}

A loose preselection of photons and $\Ppiz$ candidates is done using the
photon or $\Ppiz$ energy and the invariant $\gamma\gamma$ mass 
before the constraint to the $\Ppiz$ mass, as listed 
in table~\ref{preselcuts.table}.

\begin{table}[t]
\begin{center}
$\begin{array}{|c|}
\hline
 \multicolumn{1}{|c|}{\protect{\PDz}\ {\rm reconstruction}}
 \\
 \hline
 \hline
  1.81 \,\GeV  <   m_{\PDz}  <   1.93 \,\GeV
 \\
  0.3  <   x_{\PDz}   
 \\
  -0.85  <  \cos\thetastarpiDz  <  0.85
 \\
 \hline
 \multicolumn{1}{|c|}{
    W_{\PK}(\PK) <-1\% \ {\rm or} \ +3\% < W_{\PK}(\PK)
                     }
 \\
 \multicolumn{1}{|c|}{
  -35\% < W_{\pi}(\PK) < 0\%
                     }
 \\
 \multicolumn{1}{|c|}{
  W_{\pi}(\pi) <-1\% \ {\rm or} \ +1\% < W_{\pi}(\pi)
                     }
 \\
 \hline
 \hline
 \multicolumn{1}{|c|}{{\rm photon}\ {\rm and}\ \Ppiz\ {\rm reconstruction}}
 \\
 \hline
 \hline
 \multicolumn{1}{|c|}{
  800 \MeV < E_\gamma \ {\rm or} \ 800 \MeV < E_{\Ppiz}
                     }
 \\
 \multicolumn{1}{|c|}{
  60 \MeV < m_{\gamma\gamma} < 280 \MeV
                     }
 \\
 \hline
\end{array}$
\caption{\label{preselcuts.table}
The preselection cuts for the reconstruction of $\PDstz$ 
candidates. The selection variables are defined in the text.
}
\end{center}
\end{table}

\begin{figure}[htb]
  \begin{center}
  \unitlength 1pt
  \begin{picture}(450,492)
    \put(-71,66){\epsfig
{file=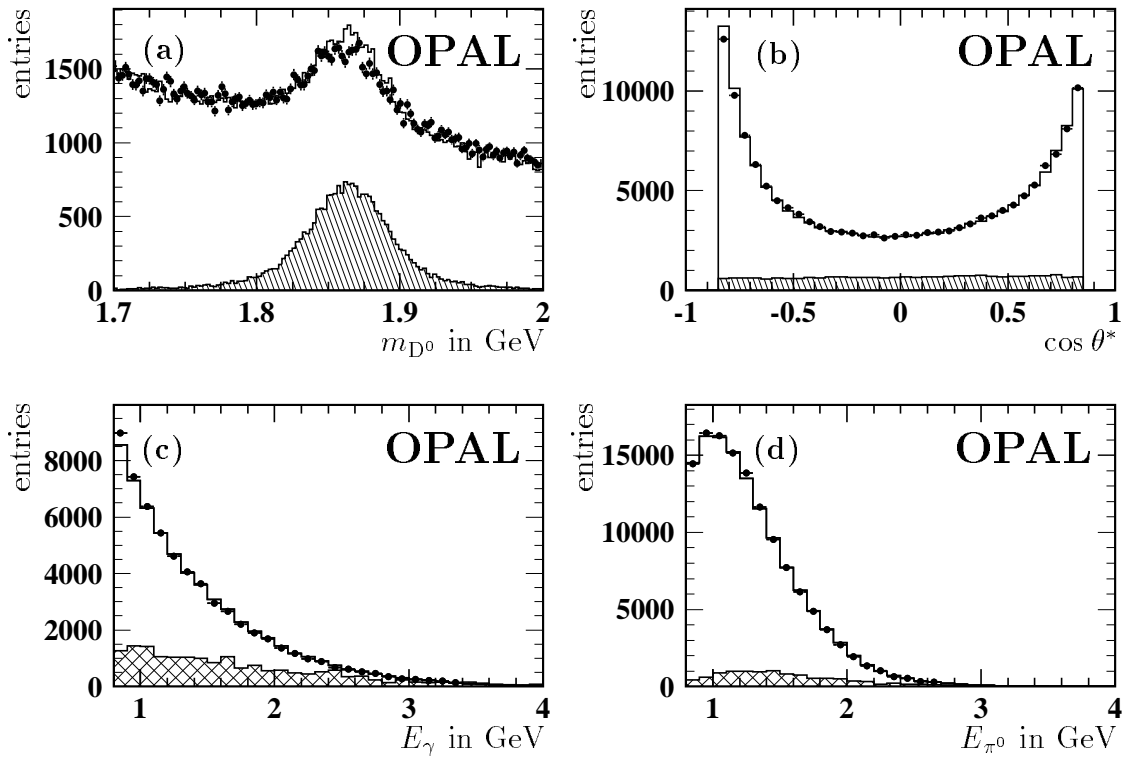,width=550pt,bbllx=0pt,bblly=347pt,bburx=550pt,bbury=792pt}}
  \end{picture}
  \caption{\label{selvars.fig}
The distributions of the four $\PDstz$ selection variables $m_{\rm D^0}$, 
$\cos\theta^*$,
$E_\gamma$ and $E_{\pi^0}$ are shown for all preselected candidates.
In each case, points with error bars correspond to the data and the open
histogram to the simulation, scaled to the same number of entries.
In figures (a) and (b), the contribution from $\PDstz$ candidates with a 
correctly reconstructed $\PDz\!\to\!\PKm\Ppip$ decay is shown as the hatched
area.
The cross-hatched areas in figures (c) and (d) correspond to 10 times the 
contribution
of correctly reconstructed photons or $\Ppiz$ mesons from a $\PDstz$ decay.
}
  \end{center}
\end{figure}

\begin{figure}[htb]
  \begin{center}\hspace{-20pt}
  \unitlength 1pt
  \begin{picture}(450,492)
    \put(-64,224){\epsfig
{file=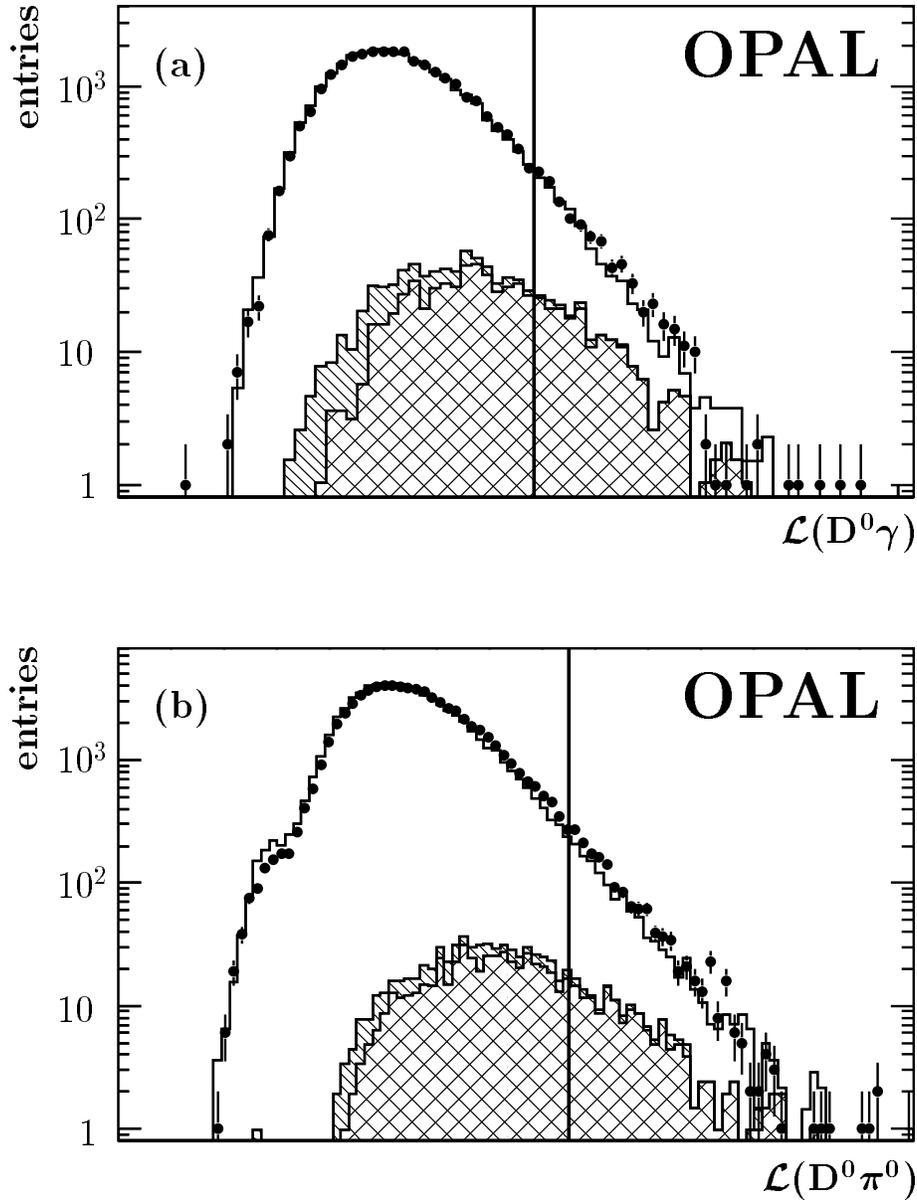,width=550pt,bbllx=0pt,bblly=405pt,bburx=550pt,bbury=792pt}}
%
%    \special{ps:gsave}
%    \put(73,415){\rotninety\rotninety\rotninety
%{\large\bf entries}}
%    \special{ps:grestore}                  
%    \special{ps:gsave}
%    \put(69,172){\rotninety\rotninety\rotninety
%{\large\bf entries}}
%    \special{ps:grestore}                  
%    \special{ps:gsave}
%    \put(346,258){\large\boldmath${\cal L}(\rm D^0 \gamma)$\unboldmath}
%    \special{ps:grestore}
%    \special{ps:gsave}
%    \put(335,13){\large\boldmath${\cal L}(\rm D^0 \pi^0)$\unboldmath}
%    \special{ps:grestore}
%    \special{ps:gsave}
%    \put(101,435){\large\bf (a)}
%    \special{ps:grestore}
%    \special{ps:gsave}
%    \put(97,192){\large\bf (b)}
%    \special{ps:grestore}
%    \special{ps:gsave}
%    \put(293,435){{\huge\bf OPAL}}
%    \special{ps:grestore}
%    \special{ps:gsave}
%    \put(289,192){{\huge\bf OPAL}}
%    \special{ps:grestore}
%    \put(53,425){\large\bf $10^3$}
%    \put(53,375){\large\bf $10^2$}
%    \put(53,325){\large\bf $10$}
%    \put(53,275){\large\bf \enspace$1$}
%    \put(53,171){\large\bf $10^3$}
%    \put(53,124){\large\bf $10^2$}
%    \put(53, 77){\large\bf $10$}
%    \put(53, 30){\large\bf \enspace$1$}
  \end{picture}
  \caption{\label{lcuts.fig}
The distributions of the likelihood functions ${\cal L}$ for the two
\protect{$\PDstz$} decay channels. Points with error bars represent the data 
and open histograms the simulation, scaled to the same number of entries.
The hatched histograms show candidates in the simulation 
with a correctly identified photon or $\Ppiz$ from a $\PDstz$ 
decay; for candidates entering the cross-hatched 
histogram, the $\PDz\!\to\!\PKm\Ppip$ decay is also correctly reconstructed. 
The vertical lines represent the cuts on the likelihood functions. No scale
is given for the likelihood values since only the relative position of the
cut is relevant.
}
  \end{center}
\end{figure}

To form $\PDstz$ candidates,
each of the preselected $\PDz$ candidates is in turn combined with every 
preselected shower photon or $\Ppiz$ candidate in the event. 
Figure~\ref{selvars.fig} shows the distribution of these $\PDstz$ candidates
in four of the most powerful selection variables.
Tighter requirements are then applied to further reduce 
the background level in the $\PDstz$ sample. They are
based on the variables listed above and on 
the total number of $\PDstz$ candidates in the 
hemisphere.
For each variable $x_i$, a purity function $\lambda_i(x_i)$ is
calculated based on the simulation as the ratio of $\c\!\to\!\PDstz$
signal candidates to all candidates:
\begin{equation}
\lambda_i(x_i) = {{\mathrm signal \over ( signal + background)}} ( x_i)
\,, 
\end{equation}
and a likelihood is constructed as 
\begin{equation}
\label{l.eqn}
{\cal L}(x_1,...,x_n) = \left(\prod_{i=1}^{n} \lambda_i(x_i) \right)
^{^{1\!} /_n} 
\,.
\end{equation}
Finally, $\PDstz$ candidates are selected using a cut on the likelihood value.
This cut has been chosen such that the statistical error on the 
result for the hadronisation fraction $\fPDstz$ 
is minimised according to an independent sample of simulated events.
The distributions of the likelihood functions ${\cal L}$ 
are shown in figure~\ref{lcuts.fig} together with the cut values.

In figure~\ref{resultd0.fig}, the
$\Delta m$=$m_{\PDstz}$$-$$m_{\PDz}$ 
mass difference 
distributions of the selected candidates are shown. The peaks at low
mass differences are due to the signal from $\PDstz$ decays.
The background shape is determined from the data as described below in 
section~\ref{backgrounddetermination.subsec}. Using this shape,
the $\PDstz$ yield is extracted from a fit to the 
mass difference distributions, which is discussed in section~\ref{fit.subsec}.

%        -------------------
\subsection{Background Determination}
%        -------------------
\label{backgrounddetermination.subsec}

Even after applying the likelihood selection, the background level in the 
samples of $\PDstz$ candidates is still high. To minimise the 
dependence on \MC{ }modelling, a method has been developed
to determine the shape of the background in the $\Delta m$
distributions from the data. 

A sample of candidates enriched in background rather than signal
is prepared with a method similar to the one for the signal 
selection (see section~\ref{dpresel.subsec}). The likelihood is 
modified to select background candidates from the $m_{\PDz}$ sideband regions
$1.70\ \GeV < m_{\PDz} < 1.84\ \GeV$ and $1.91\ \GeV < m_{\PDz} < 2.00\ \GeV$.
The sidebands are 
chosen to lie close to the signal region so that the kinematical
properties of candidates in the sidebands do not differ significantly
from those of candidates in the signal region.
The modified likelihood function is of the form
\begin{equation}
  {\cal L}'(x_1, ..., x_n)
 = 
  \left(
    \lambda_{m_{\PDz}}'(m_{\PDz}) \,
    \lambda_{\cos\thetastarpiDz}'(\cos\thetastarpiDz)
    \hspace{-10pt}
    \prod_{x_i \neq m_{\PDz}, \cos\thetastarpiDz} 
    \hspace{-15pt}
    \lambda_i(x_i) 
  \right)^{^{1\!} /_n} 
 \,,
\end{equation}
with the $\lambda$ functions left unchanged for all selection variables
except $m_{\PDz}$ and $\cos\thetastarpiDz$.
Because the original likelihood function ${\cal L}$ depends on 
$m_{\PDz}$ and $\cos\thetastarpiDz$, new functions
$\lambda_{m_{\PDz}}'(m_{\PDz})$ and 
$\lambda_{\cos\thetastarpiDz}'(\cos\thetastarpiDz)$ have to be introduced
such that the
shapes of the distributions of the likelihoods ${\cal L}$ and ${\cal L'}$
agree for background. 
In practise, this is done by constructing the $\lambda'$ functions for 
$m_{\PDz}$ and $\cos\thetastarpiDz$ separately, with the same set of values
as for the corresponding $\lambda$ function.
For the selection of background candidates, the same cut is placed
on ${\cal L}'$ as is on ${\cal L}$ for signal candidates.

One of the main requirements for the background selection is that it
correctly reproduces the $\Delta m$ shape of true background.
If the selection variables $m_{\PDz}$ and $\cos\thetastarpiDz$ are assumed to
be uncorrelated with the other selection variables and the mass difference 
$\Delta m$, the $\Delta m$ shapes of 
background candidates from the same source 
in the signal and background samples
agree by construction.
For true background, the correlations between 
$m_{\PDz}$ ($\cos\thetastarpiDz$)
and any other selection variable have been found 
in the simulation to be $\le 1\%$ ($< 10\%$),
where the largest correlations are between $\cos\thetastarpiDz$ and
$\plhmpq$ ($-9\%$) or $W_\pi(\pi)$
($-7\%$). It has been tested in the simulation that even with this level
of correlations, the background shape is correctly reproduced for background 
candidates from the same source.

Background can be classified into candidates
where a correctly reconstructed $\PDz$ is combined with a photon or $\Ppiz$
that does not come from a $\PDstz$ decay and other background,
in which the charged pion and kaon candidates do not come from the 
same $\PDz$ decay.
For these two contributions, the shapes of the $\Delta m$ distributions are 
found to be slightly different in the simulation. 
As outlined above, 
it is expected that the background determination procedure yields the correct
background shapes individually 
for both contributions. 
However, the background sample contains fewer candidates 
with a correctly reconstructed $\PDz$ than the signal 
sample.
This leads to a small bias in the overall background shape
derived from the background selection, which is 
taken into account by reweighting this shape
according to the sample composition in the signal 
and background samples as found in the simulation.

Candidates that pass the cuts on ${\cal L}$ and 
also on ${\cal L'}$ are
rejected from both the signal and the background samples.
In the simulation, it has been found that this requirement rejects 
$8\%$ of the true signal from the signal sample, and
$27\%$ of the true signal in the background sample.
It has also been verified that the fraction of candidates removed
by this requirement is the same within errors for the data and the simulation.

Although the background sample is depleted in signal candidates 
relative to the signal sample, there is a remaining $\PDstz$ signal 
contamination in the background sample. Typically, 
the signal fraction is a factor of 5 
smaller (cf.~table~\ref{efficiencies.table}) in the background samples.
These candidates have the effect of reducing the
number of signal candidates after background subtraction and are taken 
into account by calculating an effective efficiency, which is described in the 
next section.

%        -------------------
\subsection{Reconstruction Efficiency}
\label{efficiency.subsec}
%        -------------------

The efficiencies of the $\PDstz$ reconstruction are determined 
from a Monte Carlo sample which is statistically independent from the one
that has been used for the determination of the likelihood.
In this sample of $6.5$ million simulated hadronic events, the numbers of true
$\PDstz$ mesons that pass the cuts on ${\cal L}$ are computed.
The resulting efficiencies in the different decay modes 
are listed in table~\ref{efficiencies.table}, separately for $\PZz\!\to\!\cc$
and $\PZz\!\to\!\bb$ events.

As discussed in section~\ref{backgrounddetermination.subsec},
some $\PDstz$ mesons are also contained
in the background-enriched samples.  This number is determined from the 
same set of simulated events,
and an effective efficiency is computed from the values given in 
table~\ref{efficiencies.table}
that takes into account the
relative signal to background ratio of the background sample
that is given in the last column of the table.
In the fit to the mass difference distributions described in 
section~\ref{fit.subsec}, these effective efficiencies are used.

\begin{table}[htb]
\begin{center}
$\begin{array}{|@{}l@{}|@{}r@{}l||c|c|c|}
\hline
  \multicolumn{1}{|c|}{\rm decay\ chain} & 
  \multicolumn{2}{c||}{{\rm distribution}} & 
  {\rm efficiency},\ \protect{\c\!\to\!\PDst} &
  {\rm efficiency},\ \protect{\b\!\to\!\PDst} &
  (\frac{s}{b})^{\rm bkg}/(\frac{s}{b})^{\rm sig}
\\
\hline\hline
  \begin{array}{@{}r@{}l}
    \enspace
    \protect{\PDstz}\hspace{-1pt} & 
    \, \to\, \hspace{-3pt}\protect{\PDz\gamma}
   \\
    & \hspace{17.5pt}\protect{\downto\PKm\Ppip}
  \end{array}
 &
  \enspace\enspace
  \Delta m & ({\rm D^{0},\gamma}) & 
  \protect{\VeffsigDstzcgsg} &
  \protect{\VeffsigDstzbgsg}
 &
  \protect{\VetaReffsigbkgDzgamma}
\\
\hline
  \begin{array}{@{}r@{}l}
    \enspace
    \protect{\PDstz}\hspace{-1pt} & 
    \, \to\, \hspace{-3pt}\protect{\PDz\Ppiz}
   \\
    & \hspace{17.5pt}\protect{\downto\PKm\Ppip}
  \end{array}
 &
  \begin{array}{@{}r@{}}
    \Delta m \\ 
    \Delta m
  \end{array} &
  \begin{array}{@{}l@{}}
    ({\rm D^{0},\gamma}) \\ 
    ({\rm D^{0},\pi^0}) 
  \end{array} \ &
  \begin{array}{@{}c@{}}
    \protect{\VeffsigDstzcgsp} \\
    \protect{\VeffsigDstzcpsp}
  \end{array} &
  \begin{array}{@{}c@{}}
    \protect{\VeffsigDstzbgsp} \\
    \protect{\VeffsigDstzbpsp}
  \end{array}
 &
  \begin{array}{@{}c@{}}
    \protect{\VetaReffsigbkgDzgamma} \\
    \protect{\VetaReffsigbkgDzpiz}
  \end{array}
\\
\hline
\end{array}$
\caption{\label{efficiencies.table} 
The reconstruction efficiencies for a given decay chain 
in a given mass difference
distribution as determined from the simulation. 
Note that $\PDstz\!\to\!\PDz\Ppiz$ decays are measured in both the
$\Delta m(\PDz,\gamma)$ and $\Delta m(\PDz,\Ppiz)$ distributions.
The efficiencies are based on candidates with scaled energies $x_{\PDz}>0.3$.
Only statistical errors are listed.
The signal to background ratios 
$(\frac{s}{b})^{\rm bkg}$ in the background samples
relative to the ratio $(\frac{s}{b})^{\rm sig}$ in the corresponding
signal sample are given in the last column.
}
\end{center}
\end{table}

%        -------------------
\section{Measurement of \protect{\boldmath$\fPDstz$\unboldmath}}
\label{systunc.sec}
%        -------------------

The dominant sources of $\PDstz$ mesons in $\PZz$ decays are
the production in $\PZz\!\to\!\cc$ events and in bottom hadron decays.
For scaled energies $x_{\PDz}>0.3$,
the production of $\PDstz$ mesons in gluon 
splitting processes ${\rm g}\!\to\!\cc$ is highly suppressed and will be 
neglected in the following.

The hadronisation fraction $\fPDstz$ is determined in
a fit to the $\Delta m$=$m_{\PDstz}$--$m_{\PDz}$ distributions of
the selected signal and background samples.
A simultaneous fit is performed to the
$\Delta m$ distributions of both $\PDz\gamma$ and $\PDz\Ppiz$
candidates. In the fit, the contributions from b~hadron decays are
subtracted, and the fitted signals are corrected for the efficiency in
$\PZz\!\to\!\cc$ events.  In this
section, the determination of the signal contributions from b~hadron
decays, the fit to the mass difference distributions, and systematic
uncertainties of the measurement are discussed.

%        -------------------
\subsection{Subtraction of the Component from b Hadron Decays}
\label{fitresults.subsec}
%        -------------------

The determination of the contribution from $\PZz\!\to\!\bb$ events 
is based on a previous measurement of $\PDstp$ 
production~\cite{bib-Tiesdstarplus}.
Assuming isospin invariance (i.~e., equal production of $\PDstz$ and $\PDstp$ 
mesons in b~hadron decays, as well as equal hadronisation fractions 
$f(\b\!\to\!\PBm)$\,=\,$f(\b\!\to\!\PaBz)$ in $\PZz\!\to\!\bb$ decays),
the measured $\PDstp$ production rate in $\PZz\!\to\!\bb$ events is 
used as an estimate of the corresponding $\PDstz$ production rate.
This is done for scaled energies $x_{\PDz}>0.3$, and 
the remaining $\c\!\to\!\PDstz$ signal is then extrapolated to the full
$x_{\PDz}$ range.

Previously, OPAL has measured the production of $\PDstp$ mesons in 
$\PZz\!\to\!\bb$ events to be
$R_{\b} \, \mbox{\fbPDstp} \, \BrDstp \, \BrDz = 
 \VfbPDstpBr$~\cite{bib-Tiesdstarplus},
where $R_{\b}$ is the partial hadronic decay width of the $\PZz$ boson 
into $\bb$ pairs.
Using this number and the branching fraction
$\BrDstp = 0.683 \pm 0.014$~\cite{bib-PDG96},
the $\PDstp$ production rate
in $\PZz\!\to\!\bb$ events for $x_{\PDz}$$>$$0.3$ is calculated to be
\begin{equation}
 \label{betaxPDstp.eqn}
  \beta
 \equiv 
  R_\b \, \fbPDstp \, \BrDz \subxDz
 = 
  \protect{\VbetaxPDstp}
 \ .
\end{equation}
The error includes both the statistical and systematic errors.
The restriction to $x_{\PDz}$$>$$0.3$ is based on \MC{ }simulation and the
measured parameters of the b~fragmentation function.

This information is used to obtain the number of reconstructed $\PDstz$
signal candidates from b hadron decays, which is then subtracted from the
fitted signal. Thus, the hadronisation fraction $\fPDstz\subxDz$ is determined 
in the fit from
\begin{eqnarray}
 \nonumber
  R_\c \, \fPDstz \, \BrDz \subxDz
 & = & 
 \\
 \label{bsubtraction.eqn}
 & &
 \hspace{-5cm}
  \sum_{{\rm X}=\gamma,\Ppiz}
  \frac{         {\displaystyle\frac{N_{\rm sig}(\PDstz\!\to\!\PDz{\rm X})}
                                    {2 N_{\rm had}}}
         \, - \, \beta \,
                 Br(\PDstz\!\to\!\PDz{\rm X}) \
                 \varepsilon_\b^{\PDstz\to\PDz{\rm X}}
      }{         \varepsilon_\c^{\PDstz\to\PDz{\rm X}}
      }
 \,,
\end{eqnarray}
where $R_{\c}$ is defined in analogy to $R_{\b}$,
$N_{\rm sig}$ denotes the signal as obtained from the mass 
difference distribution, $N_{\rm had}$ is the number of 
hadronic $\PZz$ decays analysed, $\varepsilon_\c$ and $\varepsilon_\b$
are the effective
efficiencies for reconstructing $\PDstz$ decays with $x_{\PDz}$$>$$0.3$
from $\PZz\!\to\!\cc$ and $\PZz\!\to\!\bb$ decays, respectively, and the sum
runs over the two decay modes of the $\PDstz$. 

The assumption of equal $\PDstz$ and $\PDstp$ production in b~hadron decays
is justified from a CLEO measurement~\cite{bib-BtoDstz} of 
\begin{eqnarray}
\label{CLEOmeasurement.eqn}
  \frac{Br(\PB\!\to\!\PDstz{\rm X})}{Br(\PB\!\to\!\PDstp{\rm X})}
 =
  1.03 \pm 0.14
 \,.
\end{eqnarray}
The error on this measurement is taken into account as described in 
section~\ref{extsystunc.subsubsec}.
Decays of $\PaBz$ and $\PBm$ mesons account 
for most $\PDstz$ mesons from b~hadron decays at LEP, since
$\PaBsz$ mesons preferentially decay to final states with
a $\PDsp$~\cite{bib-PDG96}, and b~baryons are expected to decay dominantly to 
final states with a c~baryon. 

%        -------------------
\subsection{The Fit to the Mass Difference Distributions}
\label{fit.subsec}
%        -------------------

The number of reconstructed $\PDstz$ decays is determined in a 
simultaneous fit to the mass difference distributions 
of the selected signal and background samples in the two
decay modes $\PDstz\!\to\!\PDz\gamma$ and $\PDstz\!\to\!\PDz\Ppiz$.

\begin{figure}[p]
  \begin{center}
  \vspace{-32pt}
  \unitlength 1pt
  \begin{picture}(405,570)
    \put(-95,-117){\epsfig
{file=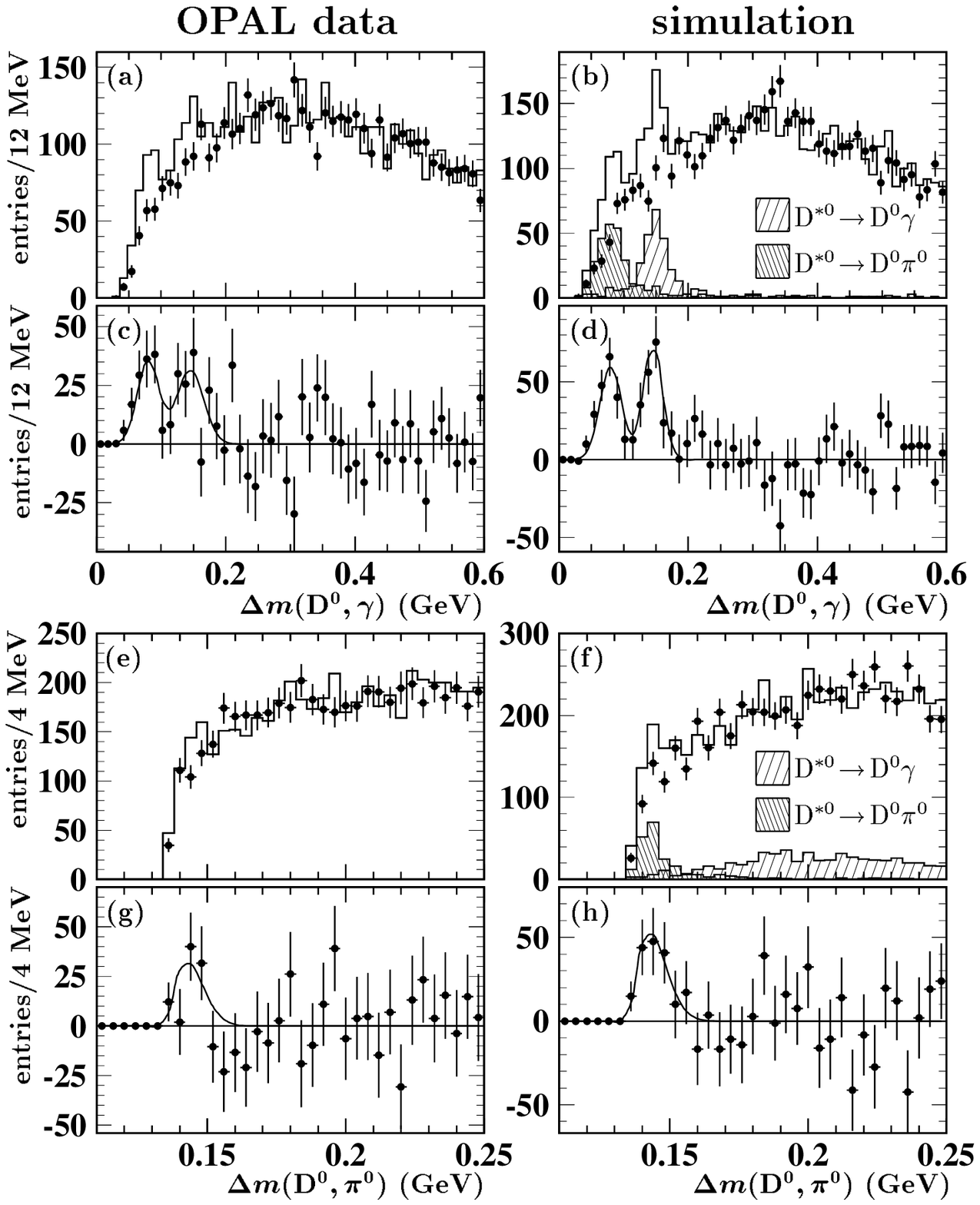,width=550pt,bbllx=0pt,bblly=42pt,bburx=550pt,bbury=784pt}}
  \end{picture}
  \caption[]{\label{resultd0.fig}
The $\protect{\Delta m(\PDz,\gamma)}$ and $\protect{\Delta m(\PDz,\Ppiz)}$ 
mass difference distributions in the data and the simulation.
\protect{\\}
In (a), (b), (e), and (f),
the distributions obtained in the signal 
selection procedure are shown as solid histograms, and those obtained in the 
background selection procedure as points with error bars, where the relative
normalisation of the latter distribution has been determined in the fit.
In addition, the hatched
histograms show the distributions of signal candidates reconstructed in 
the simulation with the signal selection procedure.
\protect{\\}
In (c), (d), (g), and (h),
the corresponding background subtracted distributions are shown
together with the fit results; the error bars show the statistical errors
only. 
}
  \end{center}
\end{figure}

Two signal peaks are visible
in the mass difference distribution of $\PDz\gamma$ candidates,
which is shown in figure~\ref{resultd0.fig}a.
The peak at the nominal $\PDstz$$-$$\PDz$ mass 
difference of 142~MeV~\cite{bib-PDG96} is due to 
$\PDstz\!\to\!\PDz\gamma$ decays, whereas the one at lower $\Delta m$
values is from $\PDstz\!\to\!\PDz\Ppiz$ decays,
where only one photon from the $\Ppiz$ decay
is combined with the $\PDz$ candidate. 

In the mass difference distribution of $\PDz\Ppiz$ candidates, the signal 
from $\PDstz\!\to\!\PDz\Ppiz$ decays
is expected around the nominal mass difference.
Additional contributions are 
expected from $\PDstz\!\to\!\PDz\gamma$ and 
$\PDstz\!\to\!\PDz\Ppiz$ decays, 
where only one of the two photons from the $\Ppiz$ decay is used. 
In both cases, an unrelated photon candidate is added
to form a $\Ppiz$ candidate. In the simulation, both contributions
have been found to be broad and very similar in shape to the 
overall background. Therefore, they are not explicitly accounted for. 

In both mass distributions, the signal is parametrised
with a functional form, while the background shape is 
determined from the background sample as decribed 
in section~\ref{backgrounddetermination.subsec},
with its normalisation determined in the fit.
A sum of two Gaussians is used as signal parametrisation for the $\Delta m$ 
distribution of $\PDz\gamma$ candidates 
(see figure~\ref{resultd0.fig}c and~d) 
in order to accomodate the contributions from 
both $\PDstz\!\to\!\PDz\gamma$ and $\PDstz\!\to\!\PDz\Ppiz$ decays.
The mean values of the Gaussians are fixed to 
$145$~MeV ($\PDstz\!\to\!\PDz\gamma$)
and $80$~MeV ($\PDstz\!\to\!\PDz\Ppiz$) as determined from the
simulation, and the widths are allowed to vary.
For $\PDz\Ppiz$ candidates, a modified Gaussian of the form
\begin{equation}
\label{sqgaus.eqn}
\frac{{\rm d}n}{{\rm d}(\Delta m)}\sim
\left(\strut\Delta m-m_{\pi^0}\right) \
\exp\left({-\frac{1}{2}\left(\frac{\Delta m
-m_{\pi^0}}{\sigma}\right)^2}\,\right)
\, ; \ \
\Delta m > m_{\Ppiz}
\end{equation}
is taken as the signal parametrisation 
(see figure~\ref{resultd0.fig}g and~h) to
account for threshold effects. 
Here, $m_{\Ppiz}$ denotes the nominal $\Ppiz$ mass~\cite{bib-PDG96}, and
the parameter $\sigma$ and the 
normalisation are determined in the fit. In the simulation, such a function 
has been found to parametrise the distribution of $\PDz\Ppiz$ 
signal candidates well.

The hadronisation fraction $\fPDstz$ is determined from a simultaneous fit to 
both $\Delta m$ distributions, with the constraint 
that the efficiency corrected number of 
$\PDstz\!\to\!\PDz\Ppiz$ decays be the same whether determined from 
the peak at lower $\Delta m$ values in the 
$\Delta m(\PDz,\gamma)$ distribution or from the 
$\Delta m(\PDz,\Ppiz)$ distribution.
Also, the efficiency corrected numbers for the two $\PDstz$ decay channels
are fixed to the world average value of the branching ratio
$Br(\PDstz\!\to\!\PDz\gamma) / Br(\PDstz\!\to\!\PDz\Ppiz) = 
0.616 \pm 0.076$~\cite{bib-PDG96}, and it is assumed that these two decay
modes saturate the $\PDstz$ width.
The contribution from bottom hadron decays, which is derived from an earlier
OPAL measurement as described in section~\ref{fitresults.subsec}, 
is subtracted in the fit.

The fit results are illustrated in figure~\ref{resultd0.fig},
where the obtained signal parametrisations are shown together with the
background subtracted mass difference distributions.
The combined fit for the $\PDstz$
measurement has a \protect{$\chi^2$} of \protect{$\VchisqDstz$}
for \protect{$\VdofDstz$} degrees of freedom.
The hadronisation fraction $\fPDstz\subxDz$ is extrapolated 
to the full range of scaled energies $x_{\PDz}$ using the simulation;
it is found to be 
\begin{equation}
  \begin{array}{rcl}
    R_\c \, \fPDstz \, \BrDz & = & \protect{\VfPDstzBrstat}
   \,,
  \end{array}
\end{equation}
where the error is statistical.

%        -------------------
\subsection{Consistency Checks}
%        -------------------
\label{syschecks.subsec}

A number of consistency checks has been performed, 
in particular to test the sensitivity of the result to 
the background subtraction procedure. Monte Carlo simulation
has been used to check that the procedure as outlined 
in section~\ref{backgrounddetermination.subsec} accurately describes the 
shape of the background in the signal sample.
Also, it has been verified that the shape is
not sensitive to the specific choice of $\lambda'$ functions.
The background determination is sensitive to correlations between 
$m_{\PDz}$ or $\cos\thetastarpiDz$ and any other selection variable.  
Any correlation seen in the data is well reproduced in the simulation.
Finally, the analysis has been repeated on the simulation,
and the generated rates are reproduced within the statistical errors. 

For all selections, the likelihood cut has been varied and the analysis 
repeated. The ranges of cut values correspond to relative
changes in efficiency of about $\pm50\%$.
The variations have been found to be consistent with
statistical fluctuations.

In the fit, the ratio $\BrDstzgammapiz$ is fixed
to its world average value of $0.616 \pm 0.076$~\cite{bib-PDG96}.
If instead, it is allowed to vary, a value of 
$\BrDstzgammapiz = \protect{\VBrgammapi0}$ 
is found, which is consistent with the world average.
The resulting hadronisation fraction $\fPDstz$ changes by
$-7.5\%$.

%        -------------------
\subsection{Systematic Uncertainties}
%        -------------------

Systematic uncertainties are related to the modelling 
of the selection variables, the detector resolution, 
the procedure used to subtract 
the background, the determination of the $\b\!\to\!\PDstz$ 
contribution to the measured 
signal, and to the extrapolation to the full range of
scaled energies $x_{\PDz}$. 
In the following sections, each of these categories is
discussed in turn. The relative values of all errors are listed in 
table~\ref{systunc.table}.

\begin{table}[htb]
\begin{center}
\begin{tabular}{|l||c|}
\hline
{\bf relative statistical error} & \protect{$\VrestDstz$} \\
\hline
\hline
\bf relative systematic errors: & \\
\enspace modelling of selection variables: & \\
\enspace \enspace $\PDz$ mass resolution 
& $\enspace1.3\%$ \\
\enspace \enspace $\langle x(\PDst) \rangle_{\PZz\to\cc\to\PDstz}$
 & $\enspace3.9\%$ \\
\enspace \enspace $\langle x({\rm X_b}) \rangle_{\PZz\to\bb\to{\rm X_b}}$
 & $\enspace3.0\%$ \\
\enspace \enspace $\PDz$ lifetime & $\enspace0.8\%$ \\
\enspace \enspace $\PDstz$ spin alignment & $\enspace0.3\%$ \\
\enspace \enspace effective shower isolation & $\enspace 9.7\%$ \\
\enspace \enspace photon pair opening angle & $\enspace2.8\%$ \\
\enspace \enspace number of $\Ppiz$ candidates & $\enspace0.2\%$ \\
\enspace \enspace number of $\PDstz$ candidates & $\enspace 1.3\%$ \\
\enspace \enspace d$E$/d$x$ & $\enspace3.6\%$ \\
\enspace \enspace d$E$/d$x$ preselection cuts & $\enspace3.1\%$ \\
\enspace \enspace fragmentation tracks & $\enspace2.7\%$ \\
\enspace \enspace shower fit & $\enspace5.9\%$ \\
\hline
\enspace detector resolution: & \\
\enspace \enspace tracking resolution
& $\enspace 6.4\%$ \\
\enspace \enspace calorimeter energy scale and resolution
& $\enspace 1.7\%$ \\
\hline
\enspace fit procedure: & \\
\enspace \enspace background normalisation & $\enspace 3.4\%$ \\
\enspace \enspace \protect{$\BrDstzgammapiz$} & $\enspace 2.0\%$ \\
\enspace \enspace background shape & $\enspace 5.5\%$ \\
\enspace \enspace contributions from other $\PDz$ decays & $\enspace 0.4\%$ \\
\enspace \enspace contributions from $\PDstp$ and $\PDsstp$ decays & $\enspace 1.4\%$ \\
\hline
\enspace b subtraction and extrapolation: & \\
\enspace \enspace subtraction of the $\b\!\to\!\PDstz$ contribution &
\protect{$\VresybsubDstz$} \\
\enspace \enspace extrapolation to \protect{$x_{\PDz}=0$} & $\enspace1.2\%$ \\
\hline
\hline
{\bf total relative systematic error} 
& \protect{$\VresyDstz$} \\
\hline
\end{tabular}
\caption{\label{systunc.table} 
A breakdown of the relative statistical and systematic errors on the 
hadronisation fraction \protect{$f(\c\!\to\!\PDstz)$}.}
\end{center}
\end{table}

%        -------------------
\subsubsection{Uncertainties from the Modelling of Selection Variables}
%        -------------------
\label{sysselproc.subsubsec}

Possible differences between distributions in data and simulation 
could influence the efficiencies and the background determination. 
These effects are 
studied separately for each selection variable. The resulting systematic errors
are assumed to be uncorrelated and are therefore added in quadrature.

Two principal methods are used to determine systematic uncertainties
in the modelling of a selection variable:\\
{\bf (A)}
Variables for which the distributions of signal candidates are
measured are treated
by reweighting the events in the Monte Carlo simulation such that for the
weighted events, the 
simulated signal distribution agrees with the measured one.
Using the likelihood given in equation~(\ref{l.eqn}), the 
background determination and the fit are then repeated and the 
original result is corrected according to the
observed difference; the 
uncertainty in the difference is interpreted as a
systematic error.
This method takes correlations between the 
selection variables into account to the 
extent that in the case of non-zero correlations, the
reweighting of events alters also the distribution of any variable that is 
correlated with the one variable in question.
This procedure is applied 
\begin{list}{$\bullet$}{\itemindent=0pt \topsep=0pt
              \parskip=0pt \parsep=0pt \itemsep=0pt 
              \partopsep=0pt}
\item
to the invariant mass distribution of $\PDz$ candidates, 
\item
to the Peterson \etal~fragmentation parameters which have been varied in the
ranges corresponding to mean scaled energies of 
$\PDstz$ mesons in $\PZz\!\to\!\cc$ decays of
$0.506 < \langle x(\PDstz) \rangle_{\PZz\to\cc\to\PDst} < 0.531$
and of weakly decaying b hadrons
in $\PZz\!\to\!\bb$ decays of 
$0.702 < \langle x({\rm X_b}) \rangle_{\PZz\to\bb\to{\rm X_b}} < 
0.730$~\cite{bib-petersonranges}, 
respectively,
\item
to the lifetime of $\PDz$ mesons, which has been varied within 
$\tau_{\PDz} = (0.415 \pm 0.004)\, {\rm ps}$~\cite{bib-PDG96}, 
\item
to the distributions of energies $E_\gamma$ and $E_{\Ppiz}$,
where the simulated events are reweighted such that the helicity angle
distributions in $\PDstz\!\to\!\PDz\gamma$ and $\PDstz\!\to\!\PDz\Ppiz$
decays agree with the measured $\PDstp$ spin alignment in $\PZz\!\to\!\cc$
events (the spin density matrix element is 
$\rho_{00} = 0.40 \pm 0.02$~\cite{bib-spinal}),
\item
to the effective separation of a shower from the closest other
shower in the electromagnetic calorimeter, 
\item
to the opening angle
of a pair of photons, and 
\item
to the numbers of reconstructed $\Ppiz$ 
and $\PDstz$ candidates. 
\end{list}
The last three of the above variables depend on the 
event topology. For these variables, possible deviations 
between data and simulation affect the signal and the background in the 
same way. The simulated events are therefore reweighted such that the
distributions of all candidates in data and simulation agree.
The analysis is then repeated, and the
observed difference in the fit result is treated as a systematic error.

\noindent {\bf (B)}
The second method is used for the
remaining selection variables, which are 
\begin{list}{$\bullet$}{\itemindent=0pt \topsep=0pt
              \parskip=0pt \parsep=0pt \itemsep=0pt 
              \partopsep=0pt}
\item
the track probabilities
calculated from d$E$/d$x$ information, 
\item
the largest longitudinal momentum of
fragmentation tracks inconsistent with being an accompanying hadron, and 
\item
the goodness of the shower fit. 
\end{list}
The corresponding function
$\lambda_i(x_i)$ is set to its maximum value for one variable $x_i$ at a time.
The resulting 
modified likelihood functions
${\cal L}_i$ and ${\cal L}_i '$ are independent of that particular variable
$x_i$, so using these
functions instead of the original likelihoods removes any possible bias
due to the simulation of $x_i$.
The selection of the signal and background $\PDstz$ candidate
samples is repeated with these modified likelihoods. By construction, these
samples also contain those candidates selected with the original likelihoods
${\cal L}$ and ${\cal L'}$. The ratio of the fit results
obtained with the original (${\cal L},{\cal L'}$) 
and the modified (${\cal L}_i,{\cal L}_i '$) likelihoods
is calculated for the data and the simulation, where the statistical 
correlation between the samples is taken into account.
The relative difference of these ratios is
interpreted as the systematic error associated with the variable $x_i$.
In the case of the signed particle identification probabilities,
the cuts for the preselection of charged kaons and pions (see 
section~\protect{\ref{dpresel.subsec}})
are retained in order to obtain a clear signal. 
In reference~\protect{\cite{bib-stableDcounting}}, an error of 
$3.1\%$ has been determined for these cuts in the
\protect{$\PDz$} selection.
This error is included as an additional systematic uncertainty.

%        -------------------
\subsubsection{Uncertainties in the Detector Resolution}
%        -------------------
\label{sysdetres.subsubsec}

\noindent{\bf Tracking resolution:}\\
Uncertainties in the modelling of the central detector are assessed by 
repeating the analysis with the tracking resolutions varied by $\pm 10\%$
around the values that describe the data best.
The redetermined efficiencies are compared with the
original ones, and 
the relative difference is interpreted as a systematic error.

\noindent{\bf Calorimeter energy scale and resolution:}\\
The energy scale and resolution of the electromagnetic calorimeter 
and the multiplicity and resolution in the time-of-flight scintillators
are treated analogously.
The reconstructed $\Ppiz$ mass distribution has been used to determine
the corresponding resolution and scaling parameters.
Thus, width and position of the $\Ppiz$ mass peak are well reproduced
in the simulation \cite{bib-sigma}, and any possible bias 
is accomodated by the variation of the detector resolution.

%        -------------------
\subsubsection{Uncertainties in the Fit Procedure}
\label{sysbkgest.subsubsec}
%        -------------------

\noindent{\bf Background normalisation:}\\
To check the determination of the background normalisation,
the fit to the mass difference distributions has been repeated with the range
restricted to values of $\Delta m(\PDz,\gamma)$$<$$0.4$ GeV and 
$\Delta m(\PDz,\Ppiz)$$<$$0.2$ GeV. Deviations from the previous results
were interpreted as systematic errors.

\noindent{\bf \boldmath$\PDstz$\unboldmath{ }branching ratio:}\\
The fit to the $\Delta m(\PDz,\gamma)$ and $\Delta m(\PDz,\Ppiz)$
distributions is constrained to the world average of the branching ratio
$\BrDstzgammapiz = 0.616 \pm 0.076$~\cite{bib-PDG96}.
The fit is repeated with this ratio varied within its errors, and the
observed difference is taken as a systematic error.

\noindent{\bf Background shape:}\\
Background candidates with a correctly reconstructed $\PDz$ lead to a 
bias in the $\Delta m$ distribution, which is taken into account in the
background subtraction as outlined in 
section~\ref{backgrounddetermination.subsec}. 
The analysis has been repeated without correcting for this bias, and half
the difference from the original result is assigned as a systematic error.

The origin of background candidates has been studied in the simulation, and
no other source for a bias of the background shape has been identified.
In particular, it has been checked that the contribution from 
photons or $\Ppiz$ mesons from $\PDstst$, $\PBst$, and $\PBstst$ decays 
is small and does not exhibit a pronounced structure.

\noindent{\bf Contributions from other \boldmath$\PDz$\unboldmath{ }decays:}\\
In the simulation, the contribution to the signal sample
of $\PDstz$ candidates from $\PDz$ mesons not decaying to $\PKm\Ppip$
was found to be $5.9\%$. This contribution is accounted for in the efficiency
determination. However, mismodelling of the branching fractions for these
other decay modes can introduce a potential bias in the $\PDstz$ measurement.
The corresponding systematic error is evaluated assuming 
relative contributions of other decay modes
as in~\cite{bib-Tiesdstarplus} and assigning errors according to the
errors on the branching fractions as given in~\cite{bib-PDG96}.

\noindent{\bf Contributions from 
\boldmath$\PDstp$ and $\PDsstp$\unboldmath{ }decays:}\\
In the reconstruction of $\PDstz$ decays, there are contributions from 
$\PDstp$ and $\PDsstp$ mesons, which can also decay via $\gamma$ or $\Ppiz$
emission. In principle, these decays can lead
to a signal in the $\Delta m$ distributions similar to that to be measured.
However, it has been found in the simulation that the relative contribution
from $\PDstp$ and $\PDsstp$ decays to the selected signal and background 
samples is approximately equal.
A small correction is applied to account for residual effects, and half
this correction is assigned as a systematic error.

It has been checked that the statistical correlation between the
$\PDstz\!\to\!\PDz\Ppiz$ signals in the $\Delta m(\PDz,\Ppiz)$ and 
$\Delta m(\PDz,\gamma)$ distributions is negligible. Also, there is only
a negligible fraction of cases where both
photons from a $\PDstz\!\to\!\PDz\Ppiz$ decay lead to an entry in the 
$\Delta m(\PDz,\gamma)$ distribution.

%        -------------------
\subsubsection{Uncertainties from the b Subtraction and the Extrapolation}
\label{extsystunc.subsubsec}
%        -------------------

\noindent{\bf Subtraction of 
the \boldmath$\b\!\to\!\PDstz$\unboldmath{ }contribution:}\\
As mentioned in section~\ref{fitresults.subsec}, it is assumed that the
production of $\PDstz$ and $\PDstp$ mesons in b~hadron decays is equal
for scaled energies $x_{\PDz}$$>$$0.3$.
The error on the production of $\PDstp$ mesons in $\PZz\!\to\!\bb$ decays
as given in equation~(\ref{betaxPDstp.eqn}) is taken into account. 
In addition, 
the production of $\PDstz$ mesons in $\PaBz$ and $\PBm$ meson decays
is varied within the range of the
CLEO measurement given in equation~(\ref{CLEOmeasurement.eqn}) to assess the
systematic error due to this assumption.
At LEP, $\PaBz$ and $\PBm$ mesons account
for $(75.6 \pm 4.4) \%$ of all weakly decaying b~hadrons~\cite{bib-PDG96}.
To evaluate the production of $\PDstz$ mesons in decays of other bottom 
hadrons, the world average of 
$Br(\PaBz\ {\rm or}\ \PBm\!\to\!\PDstp{\rm X}) 
 = (23.1 \pm 3.3)\%$~\cite{bib-PDG96} 
is compared with the recent OPAL measurement of 
$R_{\b}\fbPDstp\BrDstp\BrDz = \VfbPDstpBr$~\cite{bib-Tiesdstarplus}.
From the branching fractions as given in~\cite{bib-PDG96}, the production 
of $\PDstp$ mesons in $\PaBsz$ and $\Lambda_{\b}$ decays is then found to be
$Br(\PaBsz \ {\rm or} \ \Lambda_{\b} \!\to\! \PDstp{\rm X})
 = 0.25 \pm 0.13$. Assuming equal $\PDstz$ and $\PDstp$ production in 
$\PaBsz$ and $\Lambda_{\b}$ decays, this leads to 
$R_{\b} f(\b\!\to\!\PaBsz\ {\rm or}\ \Lambda_{\b} \!\to\! \PDstz{\rm X}) =
0.013$. A $100\%$ error is assigned to this quantity.

\noindent{\bf Extrapolation to \boldmath\protect{$x_{\PDz}=0$}\unboldmath{}:}\\
The extrapolation of the measured quantity
$R_\c \, \fPDstz \, \BrDz \subxDz$ from $x_{\PDz}$$>$$0.3$ to the full range of
scaled energies is based on the fragmentation function of Peterson
\etal~\cite{bib-peterson}.
To assess the uncertainty associated with this 
extrapolation, it is repeated with 
mean scaled energies of $\PDstz$ mesons from $\PZz\!\to\!\cc$ decays varied
in the range given in section~\ref{sysselproc.subsubsec}.
The difference from the previous result is interpreted as a systematic error.

%        -------------------
\subsection{Results of the \boldmath${\rm D}^{*0}$\unboldmath{ }Measurement}
\label{result.subsec}
%        -------------------

The production of $\PDstz$ mesons in $\PZz\!\to\!\cc$ events is 
measured to be
\begin{equation}
    \label{resultccPDsstptimesBr.eqn}
  \begin{array}{rcl}
    R_\c \, \fPDstz \, \BrDz & = & \protect{\VfPDstzBr}
   \,.
  \end{array}
\end{equation}
From this value, from the branching fraction
$\protect{\BrDz} = ( 3.83        \pm 0.12)\%$~\cite{bib-PDG96},
and from the standard model
prediction of $R_\c = 0.172$~\cite{bib-LEPEW}, the 
hadronisation fraction $\fPDstz$ is computed to be
\begin{equation}
  \begin{array}{rcl}
  f(\c\!\to\!\PDstz) & = & 
  \protect{\VfPDstz}
  .
  \end{array}
\end{equation}
Here, the last error corresponds to the error on the $\PDz$ branching fraction.

%        -------------------
\section{Measurement of \boldmath{$\PDsstp$}\unboldmath{ }Production}
\label{dsstp.sec}
%        -------------------

The decays
of $\PDsstp$ mesons are very 
simliar to those of $\PDstz$ mesons.
The techniques developed for the $\PDstz$ analysis 
can therefore be used for the reconstruction of $\PDsstp$ mesons.

The $\PDsstp$ meson decays dominantly to the $\PDsp\gamma$ final state, since 
the $\PDsp\Ppiz$ channel is suppressed by isospin invariance:
$
  Br(\PDsstp\!\to\!\PDsp\Ppiz)/
       Br(\PDsstp\!\to\!\PDsp\gamma) = 0.062^{+0.020}_{-0.018}\pm0.022
$~\cite{bib-isospinbreakingdecay}.
Thus, for the measurement of the production rate,
only the $\PDsstp\!\to\!\PDsp\gamma$ decay is used. The $\PDsp$ mesons 
are reconstructed in their decay chain 
$\PDsp\!\to\!\Pphi\Ppip$, $\Pphi\!\to\!\PKp\PKm$.
Candidates with 
invariant masses $m_\Pphi$$<$$1.05\ \GeV$ and 
$1.90\ \GeV$$<$$m_{\PDsp}$$<$$2.04\ \GeV$, 
a scaled energy of $x_{\PDsp}$$>$$0.35$, and a helicity angle of the pion
in the $\PDsp$ rest frame satisfying $-0.90$$<$$\cos\theta^*$$<$$0.95$ 
are retained if the d$E$/d$x$ requirements stated
in section~\ref{dpresel.subsec} are fulfilled for the kaon and pion candidate
tracks. The preselection of photons is identical to that described 
in section~\ref{dpresel.subsec}.

\begin{figure}[p]
  \begin{center}
  \vspace{-32pt}
  \unitlength 1pt
  \begin{picture}(405,570)
    \put(-95,-117){\epsfig
{file=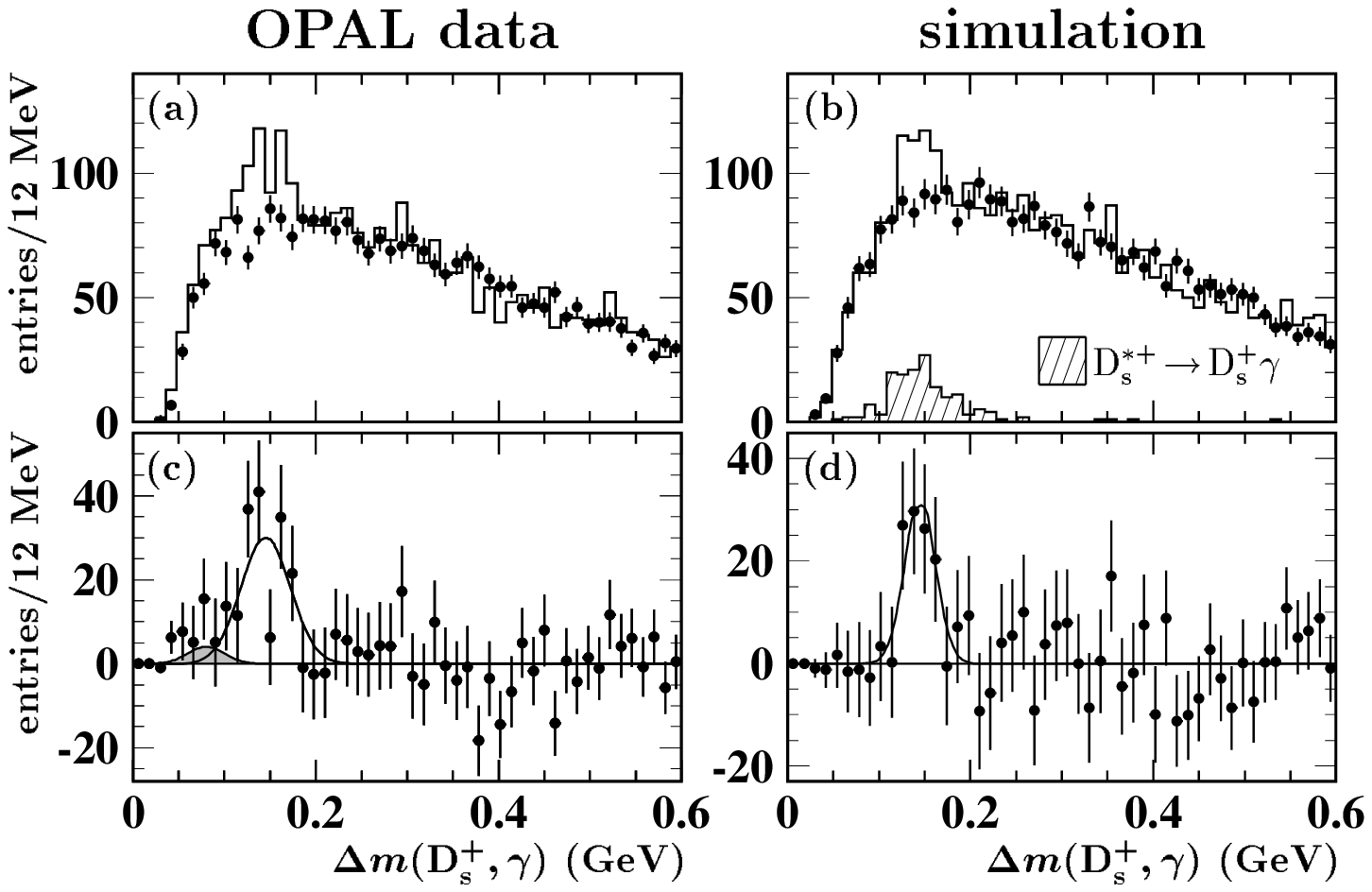,width=550pt,bbllx=0pt,bblly=42pt,bburx=550pt,bbury=784pt}}
%    \put(-24,-48){\epsfig
%{file=fig/resultdsp.fig2,width=450pt,bbllx=0pt,bblly=42pt,bburx=550pt,bbury=784pt}}
%    \special{ps:gsave}
%    \put(70,508){\Large\bf OPAL data}
%    \special{ps:grestore}                  
%    \special{ps:gsave}
%    \put(272,508){\Large\bf simulation}
%    \special{ps:grestore}                  
%    \special{ps:gsave}
%    \put(-1.5,408){\rotninety\rotninety\rotninety
%{\bf entries/12 MeV}}
%    \special{ps:grestore}                  
%    \special{ps:gsave}
%    \put(28,486){\bf (a)}
%    \special{ps:grestore}
%    \special{ps:gsave}
%    \put(225,486){\bf (b)}
%    \special{ps:grestore}
%%
%    \special{ps:gsave}
%    \put(-13,298){\rotninety\rotninety\rotninety
%{\bf entries/12 MeV}}
%    \special{ps:grestore}                  
%    \special{ps:gsave}
%    \put(73,256){\boldmath$\Delta m(\PDsp,\gamma)$\unboldmath {\bf{ }(GeV)}}
%    \special{ps:grestore}
%    \special{ps:gsave}
%    \put(269,256){\boldmath$\Delta m(\PDsp,\gamma)$\unboldmath {\bf{ }(GeV)}}
%    \special{ps:grestore}
%    \special{ps:gsave}
%    \put(8,376){\bf (c)}
%    \special{ps:grestore}
%    \special{ps:gsave}
%    \put(205,376){\bf (d)}
%    \special{ps:grestore}
%%
%    \put(294,408){$\PDsstp\!\to\!\PDsp\gamma$}
  \end{picture}
\vspace{-230pt}
  \caption[]{\label{resultdsp.fig}
The mass difference distribution for the decay $\PDsstp\!\to\!\PDsp\gamma$.
As before, in (a) and (b), the solid histogram corresponds to the signal 
sample, and
points with error bars to the background sample. The signal contribution
in the simulation is indicated by the hatched histogram.\protect{\\}
The corresponding background subtracted distributions are shown in (c) and
(d) together with the fitted signal parametrisation. The shaded Gaussian
in figure (c) shows the expected contribution from $\PDsstp\!\to\!\PDsp\Ppiz$
decays which is fixed in the fit.
}
  \end{center}
\end{figure}

The $\PDsstp$ candidates are selected with a cut on a likelihood 
using the variables described in section~\ref{dpresel.subsec} accordingly, and 
taking the reconstructed mass of the $\Pphi$ and the cosine of the 
angle between the $\PDsp$ and one of the kaons
in the $\Pphi$ rest frame as additional inputs.
The background shape is determined from candidates in sidebands of 
$m_\Pphi$ and $m_{\PDsp}$
using the technique described in section~\ref{backgrounddetermination.subsec}.
The resulting mass difference distribution is shown in 
figure~\ref{resultdsp.fig}.

No attempt is made to separate the contributions from the processes
$\c\!\to\!\PDsstp$ and $\b\!\to\!\PDsstp$.
However, the efficiencies for $\PDsstp$ reconstruction in $\PZz\!\to\!\cc$ and
$\PZz\!\to\!\bb$ events have been found to be equal within errors. Thus, 
it is still reasonable to extract
the overall $\PDsstp$ production in $\PZz$ decays from a fit to the
$\Delta m(\PDsp,\gamma)$ distribution.
No isospin violating decays $\PDsstp\!\to\!\PDsp\Ppiz$ have been 
generated in the simulation; in the data, a small contribution from 
such decays is expected at $\Delta m$ values below the nominal
$\PDsstp$--$\PDsp$ mass difference as shown in figure~\ref{resultdsp.fig}c. 
A double Gaussian is used to parametrise the signal, where the Gaussian at
lower $\Delta m$ is fixed to the expectation 
and varied by $\pm100\%$ in case of the data.
The number of $\PDsstp$ mesons per hadronic $\PZz$ decay with 
$x_{\PDsp}$$>$$0.35$ is found to be
\begin{equation}
\label{nbarxPDsstp.eqn}
  \begin{array}{r}
  \bar{n}(\PZz\!\to\!\PDsstp) 
\, Br(\PDsstp\!\to\!\PDsp\gamma)
  \, \protect{\BrDsp} \, \protect{\BrPhi} \subxDsp \hspace{2cm} 
 \\
  \hspace{5cm} = 
  \protect{\VnbarxPDsstpBr}
 \,, 
  \end{array}
\end{equation}
where the first error is statistical and the second systematic. 
The fit to the data has a $\chi^2$ of \protect{$\VdatachisqDspgamma$}
for \protect{$\VdofDspgamma$} degrees of freedom.

The contributions to the systematic error were evaluated in a similar way as
described earlier. 
Table~\ref{systuncdsstp.table} lists the most important errors. 
The largest systematic error on the value in equation~(\ref{nbarxPDsstp.eqn})
is introduced from the uncertainty in the
background shape. The background sample is
taken from a sideband in $m_{\Pphi}$, and a 
correction has to be applied to account for the
fact that fewer correctly reconstructed $\Pphi\!\to\!\PKp\PKm$ decays
enter the background than the signal sample.
When the fit is repeated with the same signal parametrisation
plus an exponential 
to account for potential problems in the modelling of this bias, a consistent
result is found. The deviation from the previous result is interpreted as
systematic error. 
No other sources have been identified that could lead
to a significant bias in the background shape.
\begin{table}[htb]
\begin{center}
\begin{tabular}{|l||c|}
\hline
{\bf relative statistical error} & \protect{$\VrestDsstp$} \\
\hline
\hline
\bf relative systematic errors: & \\
\enspace extrapolation to $x_{\PDsp}$=0 & ($19.6\%$) \\
\enspace background shape & $11.8\%$ \\
\enspace tracking resolution & $\enspace8.5\%$ \\
\enspace $\PDsp$ lifetime & $\enspace8.5\%$ \\
\enspace effective shower isolation & $\enspace8.1\%$ \\
\enspace shower fit & $\enspace6.9\%$ \\
\enspace d$E$/d$x$ preselection cuts & $\enspace6.4\%$ \\
\enspace $\langle x({\rm X_b}) \rangle_{\PZz\to\bb\to{\rm X_b}}$
 & $\enspace6.2\%$ \\
\enspace others (evaluation similar to the $\PDstz$) & $\enspace9.5\%$ \\
\hline
\hline
{\bf total relative systematic error} 
& $\begin{array}{@{}c@{}}
     \protect{\VresyxDsstp} \\
     (\protect{\VresyDsstp})
   \end{array}$ \\
\hline
\end{tabular}
\caption{\label{systuncdsstp.table} 
A breakdown of the relative statistical and dominant systematic errors on the 
$\PDsstp$ measurement. The errors in brackets apply only when considering
the value which has been extrapolated to all scaled energies $x_{\PDsp}$.
}
\end{center}
\end{table}

The above $\PDsstp$ measurement 
is extrapolated to the full range of scaled energies $x_{\PDsp}$. Here, a large
systematic uncertainty is introduced from the unknown relative contribution
from $\c\!\to\!\PDsstp$ and $\b\!\to\!\PDsstp$ components.
When varying the unknown ratio $\fbPDsstp/\fbPDsp$ within $0.6\pm0.2$, 
one finds the total rate
\begin{equation}
  \begin{array}{r}
  \bar{n}(\PZz\!\to\!\PDsstp) \, Br(\PDsstp\!\to\!\PDsp\gamma)
  \, \protect{\BrDsp} \, \protect{\BrPhi} \hspace{2cm} 
 \\
  \hspace{5cm} = 
  \protect{\VnbarPDsstpBr}
 \,, 
  \end{array}
\end{equation}
where the third error is introduced from the extrapolation to the full range 
of scaled energies.
This rate is consistent with expectations and can be regarded as a 
further cross-check of the $\PDstz$ analysis,
which uses very similar techniques.

%        -------------------
\section{The Relative Production Rate of Charmed Vector Mesons in 
         \boldmath${\rm Z}^0\!\to\!{\rm c\bar{c}}$\unboldmath{ }Decays}
%        -------------------
\label{vvp.sec}

The relative production rate of vector mesons containing
the primary quark, 
$\VVP$, is an important parameter in fragmentation and hadronisation models. 
Hadrons 
that contain the primary quark can in principle be produced either directly or 
in decays of higher resonances. Therefore, there are two possible definitions
of $\VVP$: the quantity $\VVPF$, where inclusive production 
(including decays of higher resonances) is considered, and $\VVPP$, 
which is defined when considering direct production only. These two 
values may differ due to the effects of decays of higher resonances,
where some decays are forbidden by spin and parity conservation.

In previous investigations of $\VVP$ in the charm 
system~\cite{bib-christoph,bib-VVPvalues,bib-rostock},
isospin invariance between the neutral and charged non-strange
vector mesons was assumed. Given the measurement 
of the hadronisation fraction $\fPDstz$ presented in the 
first part of this paper, an explicit check of 
this assumption is now possible. 

Recently, a measurement
of the spin alignment of $\PDstp$ mesons in $\PZz\!\to\!\cc$ decays has 
been presented~\cite{bib-spinal}. While $\VVP$ denotes the relative
production of vector and pseudoscalar mesons, spin alignment measurements
provide information on the relative production of different vector meson
spin states.
Thus, the combination of these two measurements
provides further insight into the inclusive production 
of charmed mesons in the fragmentation process~\cite{bib-christoph}. 

In the first part of this section, the existing OPAL 
measurements of the production of charmed mesons with no orbital angular 
momentum are used to derive a value of $\VVPF$, which is 
interpreted in conjunction with the $\PDstp$ spin alignment.

In the second part, additional input from 
excited $\PD$ states is used to derive $\VVPP$.
This quantity can be compared more directly to model calculations, 
since such 
models generally make predictions for primary hadron production. However, 
a determination of $\VVPP$ is experimentally challenging because of 
the difficulties in assessing the fraction of hadrons produced in decays
of higher resonances.
In this paper, the production of $L$=1 mesons is taken into account, while
production and decay of higher states is considered a part of the 
fragmentation process.
For each of the light flavours u, d, and s,
four charmed mesons with orbital angular momentum $L$=1 are predicted.
The naming convention is ${\rm D}_J^{(*)}$, where $J$ denotes the total spin 
of the meson, and an asterisk indicates that the meson has parity $(-1)^J$.
In the following, these are collectively referred to as $\PDstst$ mesons.
Using the recent OPAL measurements of the hadronisation fractions 
$\fPDststCz$, $\fPDststDz$, and $\fPDsststCp$~\cite{bib-Ddoublestar}, 
decays of $\PDstst$ mesons can be taken into account to study the 
$\VVPP$ ratio.
First, the dependence of $\VVPP$ on the production of the 
unmeasured $L$=$1$ resonances is discussed, and a model independent 
formula for the calculation of $\VVPP$ is derived.
Second, the production of the unmeasured $L$=$1$ resonances is assessed in 
a simple spin counting picture.

The following discussions are based on the $\fPDstz$ measurement
described in the first part of this paper
and on previous OPAL measurements of charmed meson 
hadronisation fractions.
Table~\ref{vvptable.table} contains an overview of the values entering
the following computations.

\begin{table}[htb]
\[
\begin{array}{|l|l|}
\hline
 \multicolumn{1}{|c|}{\rm charmed\ meson\ production,\ OPAL\ measurements} & 
 \multicolumn{1}{c|}{\rm value\ and\ reference} \\
\hline
 R_\c \ \fPDstz \ \BrDz &
 \protect{\VfPDstzBrfortable} \ \
 \protect{\rm (section~\ref{result.subsec})} 
\\
 R_\c \ \fPDstp \ \BrDstp \, \BrDz \! &
 (1.041 \pm 0.020 \pm 0.040) \times 10^{-3} \ \
 \protect{\rm \cite{bib-Tiesdstarplus}} 
\\
 R_\c \ \fPDz \ \BrDz &
 (0.389 \pm 0.027 ^{+0.026}_{-0.024}) \times 10^{-2} \ \
 \protect{\rm \cite{bib-stableDcounting}} 
\\
 R_\c \ \fPDp \ \BrDp &
 (0.358 \pm 0.046 ^{+0.025}_{-0.031}) \times 10^{-2} \ \
 \protect{\rm \cite{bib-stableDcounting}} 
\\
 R_\c \ f(\c\!\to\!{\rm D_1^0 \ or \ D_2^{*0}}) \ 
     Br({\rm D_1^0 \ or \ D_2^{*0}}\!\to\!\PDstp\Ppim) &
 (4.2 \pm 1.1 ^{+0.5}_{-0.7} {}^{+0.2}_{-0.3}) \times 10^{-3} \ \
 \protect{\rm \cite{bib-Ddoublestar}} 
\\
  \frac{f(\c\to{\rm D_1^0}\to\PDstp\Ppim)}
       {f(\c\to{\rm D_1^0}\to\PDstp\Ppim) + f(\c\to\PDststDz\to\PDstp\Ppim)}
 & 0.56 \pm 0.15 ^{+0.03}_{-0.04} \ \
 \protect{\rm \cite{bib-Ddoublestar}} 
\\
 R_\c \ f(\c\!\to\!{\rm D_{s1}^+}) &
 (2.8 ^{+0.8}_{-0.7} \pm 0.3 \pm 0.4) \times 10^{-3} \ \
 \protect{\rm \cite{bib-Ddoublestar}} 
\\
\hline
\hline
 \multicolumn{1}{|c|}{\rm branching\ fractions\ and\ ratios} & 
 \multicolumn{1}{c|}{\rm value\ and\ reference} \\
\hline
 \protect{\BrDz} &
 0.0383 \pm 0.0012 \ \
 \protect{\rm \cite{bib-PDG96}}
\\
 \protect{\BrDp}/\protect{\BrDz} &
 2.35 \pm 0.16 \pm 0.16 \ \
 \protect{\rm \cite{bib-BrDpDz}}
\\
 \protect{\BrDstp} &
 0.683 \pm 0.014 \ \
 \protect{\rm \cite{bib-PDG96}}
\\
 \frac{Br({\rm D}_2^*\to\PDst\pi)}
 {Br({\rm D}_2^*\to\PDst\pi) + Br({\rm D}_2^*\to\PD\pi)}
 \approx Br({\rm D}_2^*\!\to\!\PDst\pi) &
 0.311 \pm 0.051 \ \
 \protect{\rm \cite{bib-PDG96}} 
\\
 \frac{Br({{\rm D_{s2}^{*+}}\to\PDst\PK})}
 {Br({{\rm D_{s2}^{*+}}\to\PDst\PK}) + Br({{\rm D_{s2}^{*+}}\to\PD\PK})}
 \approx Br({{\rm D_{s2}^{*+}}\!\to\!\PDst\PK}) &
 0.107 \pm 0.016 \ \
 \protect{\rm \cite{bib-Ddoublestar,bib-theorBrsdoublestar,bib-CLEOlimit}} 
\\
\hline
\hline
 \multicolumn{2}{|c|}{\rm assumption\ on\ relative\ \protect{\PDstst}\ meson\ 
                          production\ in\ a\ spin\ counting\ model} \\
\hline
 \begin{array}{@{}c@{\,:\,}c@{\,:\,}c@{\,:\,}c@{}}
   \protect{\fPDststAz} & \protect{\fPDststBz} & 
   \protect{\fPDststCz} & \protect{\fPDststDz} 
   \\
   \protect{\fPDsststAp} & \protect{\fPDsststBp} & 
   \protect{\fPDsststCp} & \protect{\fPDsststDp} 
 \end{array}
 &
 \left.
 \!
 \begin{array}{@{}l@{}}
   \protect{\phantom{:\!\!}}
   \\
   \protect{\phantom{:\!\!}}
 \end{array}
 \right\}
 \hspace{5pt}
  1 : 3 : 3 : 5 
\\
\hline
\end{array}
\]
\caption{\label{vvptable.table} 
  The measurements and assumptions entering the computation of $\VVP$ values.
  }
\end{table}

%        -------------------
\subsection{Tests of Isospin Invariance}
%        -------------------
\label{isospin.subsec}

Isospin invariance suggests equal primary production rates for corresponding
$\c\bar{\u}$ and $\c\bar{\d}$ mesons. Thus, the hadronisation
fractions $\fPDstz$ and $\fPDstp$ are expected to be the same, as long as
decays of higher resonances contribute equally to both $\PDstz$ and $\PDstp$
production.
For the vector mesons $\PDstz$ and $\PDstp$, isospin invariance can 
therefore be tested directly 
from the $\PDstz$ and $\PDstp$ production rate measurements and the 
branching fraction $\BrDstp$ as listed in table~\ref{vvptable.table},
resulting in a ratio of
\begin{equation}
\frac{f(\c\!\to\!\PDstz)}{f(\c\!\to\!\PDstp)}=
\protect{\VfPDstzPDstp}
\,,
\end{equation}
consistent with $1$.

Since $\PDstz\!\to\!\PDp\Ppim$ decays are kinematically forbidden
while $\PDstp\!\to\!\PDz\Ppip$ decays are not,
the observed hadronisation fractions of $\PDz$ and $\PDp$ mesons are
expected to differ even if the primary 
production rates are equal.  When taking into account these $\PDstp$ decays
and assuming 
isospin invariance between $\PDz$ and $\PDp$ mesons, the ratio
\begin{equation}
\label{isoconstraint.eqn}
  R
 \equiv
  \frac{ \displaystyle \fPDz - \fPDp }
       { \displaystyle \fPDstz + \left( 2 \BrDstp - 1 \right) \fPDstp }
 \,
\end{equation}
is expected to be equal to one. 
From the values given in table~\ref{vvptable.table}, one obtains
$R = \protect{\VRjso}$,
consistent with~$1$.
Under the assumption of equal hadronisation fractions $\fPDstz=\fPDstp$, 
other experiments~\cite{bib-VVPvalues} have obtained values 
consistent with this result.

With isospin invariance thus confirmed within experimental errors, 
the hadronisation fractions of $\PDstz$ and $\PDstp$ 
mesons are assumed to be equal in the following.
The mean $\PDstzp$ hadronisation fraction is then defined as the weighted 
average of the values of $\fPDstzp$ determined from 
equation~(\ref{isoconstraint.eqn}) and from the direct measurements of 
$\fPDstz$ and $\fPDstp$. The resulting value of 
\begin{equation}
\label{weightedDstar.eqn}
R_\c \, \fPDstzp \, \BrDstp \, \BrDz = \protect{\VfPDstBr}
\end{equation}
is used in the computation of $\VVP$ in the following sections.

%        -------------------
\subsection{Effective Charmed Meson Production}
%        -------------------
\label{vvpf.subsec}

The effective value $\VVPF$ is calculated from 
the mean $\PDstzp$ hadronisation fraction derived in the previous 
section and 
the measurements listed in table~\ref{vvptable.table} to be
\begin{eqnarray}
  \label{vvpf.eqn}
  \protect{\VVPF} = 
  \frac{ 2 f(\c\!\to\!\PDstzp) }
       { f(\c\!\to\!\PDz)   + f(\c\!\to\!\PDp) } =
  \VVVPF
  \,.
\end{eqnarray}
Here, the assumption has been made that isospin invariance is valid, as
tested in the previous section. If, instead,
the measured hadronisation fractions 
are directly combined without the assumption of isospin invariance for
$\PDstz$ and $\PDstp$ production, a value of
\protect{$\VVPF =   \frac{ f(\c\to\PDstz) + f(\c\to\PDstp) }
                         { f(\c\to\PDz)   + f(\c\to\PDp) } =
          \VVVPO$}
is obtained.
In principle, the measurements are expected to be 
correlated. The largest correlation is expected between
$\fPDz$ and $\fPDstp$ and has been estimated to be smaller 
than $30\%$. This introduces a systematic error on $\VVPF$ 
of less than $0.01$.

The result from equation~(\ref{vvpf.eqn}) can be interpreted in connection
with the $\PDst$ spin alignment.
In the fragmentation process, four different spin states of 
charmed mesons without orbital excitation 
can be formed: The vector mesons $\PDstz$, $\PDstp$, and
$\PDsstp$ with states $J=1$ and $m=-1, 0, +1$, 
as well as the pseudoscalar
mesons $\PDz$, $\PDp$, and $\PDsp$ with $J=m=0$. 
The relative inclusive production probabilities ${\cal P}$ for these states
are related to $\VVPF$ and to the spin density matrix element $\rho_{00}$
of $\PDst$ mesons from $\PZz\!\to\!\cc$ events
via~\cite{bib-christoph,bib-falkpeskin}
\begin{eqnarray}
  \label{relprob.eqn}
  \begin{array}{rcl}
    \protect{\VVPF}
   & = &
    {\cal P}_{J=1}^{m=0}+{\cal P}_{J=1}^{m=\pm1}
   \ \ \ {\rm and}
   \\
    \rho_{00}
   & = & 
    {\cal P}_{J=1}^{m=0}/({\cal P}_{J=1}^{m=0}+{\cal P}_{J=1}^{m=\pm1})
   \ .
   \rule{0pt}{2.6ex}
  \end{array}
\end{eqnarray}
Here, ${\cal P}_{J=1}^{m=\pm1}$ denotes the
sum of the production probabilities for the $m=+1$ and $m=-1$ states that 
cannot be distinguished experimentally.
The spin density matrix element $\rho_{00}$ gives the probability to find
a vector meson in the $m=0$ state.
A simple spin counting model suggests values of 
${\cal P}_{J=1}^{m=\pm1}=\frac{1}{2}$ and
${\cal P}_{J=1}^{m=0}= {\cal P}_{J=0}^{m=0}= \frac{1}{4}$ (see for 
instance~\cite{bib-falkpeskin}). 

For $\PDstp$ mesons from $\PZz\!\to\!\cc$ decays
with scaled energies $x_{\PDstp}$$>$$0.2$,
$\rho_{00}$ has been measured at OPAL to 
be $\rho_{00}= 0.40 \pm 0.02$~\cite{bib-spinal}.
From this result and the above value of $\VVPF$, the
production probabilities
\begin{eqnarray}
\nonumber
{\cal P}_{J=1}^{m=0}    = \rho_{00}\VVPF     & = & \protect{\VspinalPA}
\,,\\
{\cal P}_{J=1}^{m=\pm1} = (1-\rho_{00})\VVPF & = & \protect{\VspinalPB}
\,,\enspace{\rm and}\\
\nonumber
{\cal P}_{J=0}^{m=0}    = 1-\VVPF            & = & \protect{\VspinalPC}
\end{eqnarray}
can be derived,
where it has been assumed that the measured spin density matrix 
element $\rho_{00}$ applies to both $\PDstz$ and $\PDstp$ mesons in 
$\PZz\!\to\!\cc$ events. 

These probabilities show a clear deviation from the simple 
spin counting picture. The production of vector mesons is suppressed
in favour of pseudoscalar mesons. This
is mostly due to a suppression of the $m$=$\pm1$ vector states, 
while the production of the $J$=$1$, $m$=$0$
state agrees within errors with a spin counting picture.

An overall suppression of vector meson production is, for instance,
expected in thermodynamic models, where the constituents of hadrons
are pictured as a gas with a temperature $T$, 
such that a relative suppression of the heavier states by a factor of 
$\exp(-\Delta m/T)$ is predicted.
From the above value of $\VVPF$, 
the temperature is calculated to be $T=\VT$, consistent
with the value determined in~\cite{bib-becattini}. 
It should be noted, however, that
thermodynamic models fail to explain the observed non-zero spin 
alignment.

%        -------------------
\subsection{Primary Charmed Meson Production}
%        -------------------
\label{vvpp.subsec}

For the determination of $\VVPP$,
the effects of the decays of $L$=$1$ charmed mesons have to be taken into
account. The hadronisation 
fractions $\fPDststCz$, $\fPDststDz$, and $\fPDsststCp$ have been measured
at LEP, whereas the production of the broad
resonances $\PDststA$ and $\PDststB$ has not yet been measured.
Since the charged $\PDststp$ mesons 
have not yet been observed in $\PZz$ decays,
isospin invariance is assumed to be valid in charm fragmentation 
to assess the production of these resonances, 
which implies equal hadronisation fractions
for corresponding $\PDststp$ and $\PDststz$ states.
Furthermore, it is assumed in the following that for $\PDstst$ mesons, the 
relative production of the different spin states does not depend on the
light quark flavour. 

The ratio $\VVPP$ can be expressed as a function of 
two unmeasured hadronisation fractions,
$\fPDststAz$ and $\fPDststBz$, and two known
parameters $A$ and $B$ which depend on the 
measurements listed in table~\ref{vvptable.table}:
\begin{equation}
\label{vvpi.eqn}
\VVPP = 
\frac{ A - R_\c f(\c\!\to\!\PDststBz) }
     { B - R_\c \left[ f(\c\!\to\!\PDststAz) + f(\c\!\to\!\PDststBz) \right] }
\,.
\end{equation}
The complete derivation and the exact formulae for $A$ and $B$ are 
given in appendix A. 
The parameters $A$ and $B$ are found to be
\begin{equation}
  \begin{array}{lcl}
    A 
   & = &
    \protect{\VA} 
    \enspace{\rm and}
   \\[0.25ex]
   \vspace{0.25ex}
    B
   & = & 
    \protect{\VB} 
    \,,
   \end{array}
\end{equation}
with a positive correlation between $A$ and $B$ of 
$\protect{\VABcorrel}$.

\begin{figure}[htb]
  \begin{center}
  \unitlength 1pt
  \begin{picture}(450,410)
    \put(-71,182){\epsfig
{file=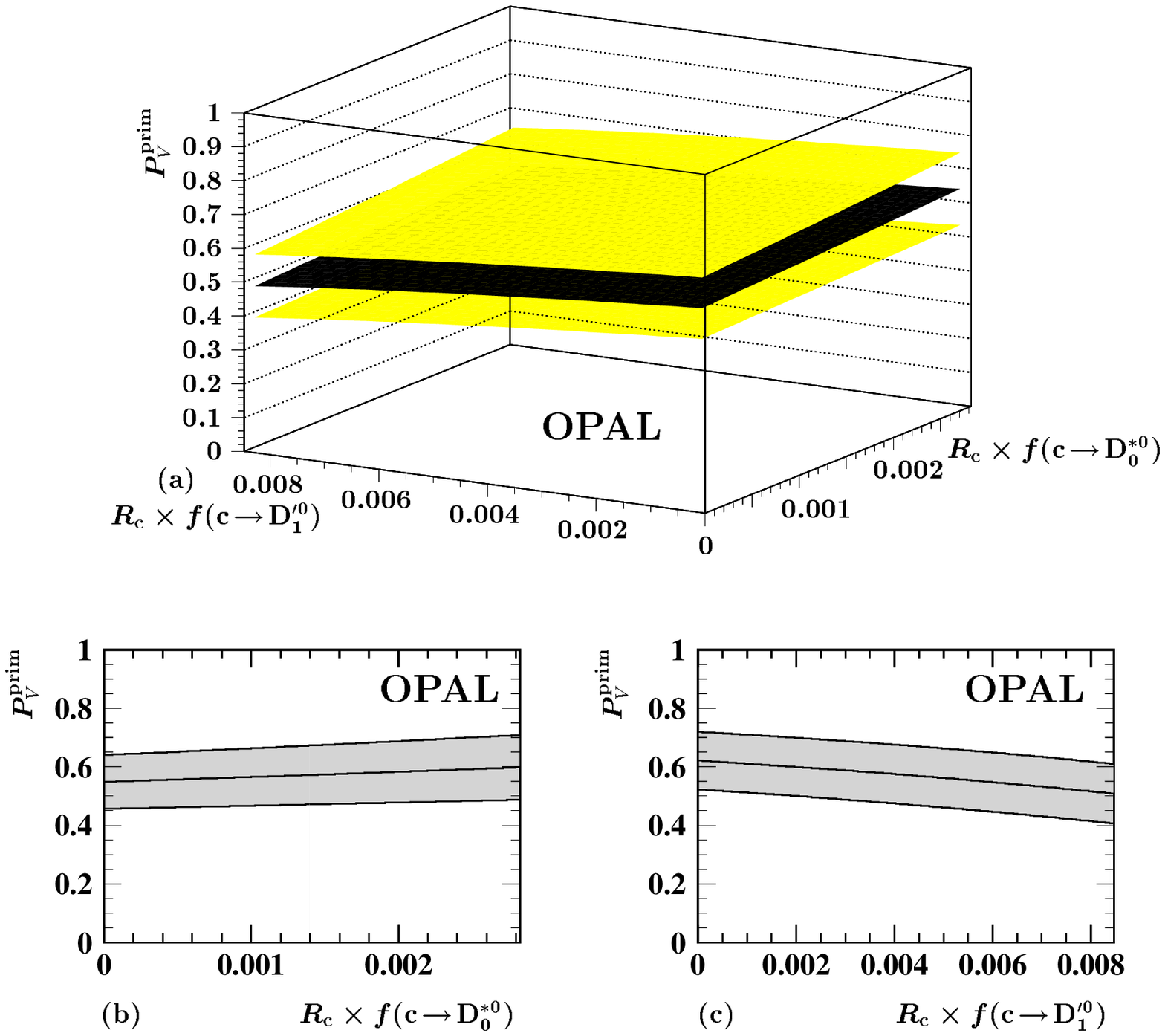,width=550pt,bbllx=0pt,bblly=405pt,bburx=550pt,bbury=792pt}}
  \end{picture}
  \caption{\label{vvppdependence.fig}
The dependence of $\protect{\VVPP}$ on the two unmeasured multiplicities
$\protect{R_\c \times \fPDststAz}$ and $\protect{R_\c \times \fPDststBz}$ 
is shown in figure (a).  Each of these quantities is varied between 0 and 
twice the value as expected from a spin counting picture.
Plots (b) and (c) show the dependence on each
one of these multiplicities when the other
is fixed at the spin counting expectation. The error contours correspond to 
one standard deviation.  They
include the experimental error on the relative production of
$\protect{\PDststCz}$ and $\protect{\PDststDz}$ mesons as measured
in~\protect{\cite{bib-Ddoublestar}}.
}
  \end{center}
\end{figure}

In figure~\ref{vvppdependence.fig}, the results for $\VVPP$ are
shown as a function of the two unknown hadronisation fractions.
In general, low values of $\VVPP$ are obtained for small $\PDststA$ and
large $\PDststB$ production and vice versa. 
This analysis shows that the value of $\VVPP$ is not
very sensitive to the production of the unmeasured broad $\PDstst$ resonances.
Generally, the range of $\VVPP$ values is consistent with predictions.
However, to test a given model, a clearer statement can be made when
following the procedure outlined below.

In contrast to the above discussion, a test of any specific fragmentation model
can be performed when using its prediction for the relative 
primary production of the different $L$=0 and $L$=1 states. 
While the prediction for the $L$=0 states is directly equivalent to
a prediction of $\VVPP$, the
model prediction in the $L$=1 sector provides information on the production
of the unmeasured $L$=1 resonances, which can in turn be used in conjunction
with the measured hadronisation fractions to obtain an experimental value
of $\VVPP$. Such a test is described in the following for the simplest case,
a spin counting model.

A spin counting model predicts the relative primary production of $\PD$ and
$\PDst$ mesons to be $1:3$. At the same time, for $\PDstst$ mesons,
the relative production of the $\PDststA:\PDststB:\PDststC:\PDststD$ 
resonances is predicted to be $1:3:3:5$. When using the latter prediction, 
$\VVPP$ is found to be
\begin{equation}
\label{VVPPspincounting.eqn}
  \protect{\VVPP}({\rm spin\ counting\ for\ }\PDstst) 
 = 
  \protect{\VVVPPnospincounting}
 \,.
\end{equation}
However, since the validity of spin counting was assumed in the calculation, 
the value of $\VVPP$ is fixed and should be $0.75$.
Thus, from the discrepancy between the above value and $0.75$, 
the simultanous description of both $L$=0 and $L$=1 charmed meson production
in a spin counting picture is disfavoured 
by $\VnsigmaVVPP$ 
standard deviations.
Within the framework of this model test, the relative production
of the four $\PDstst$ spin states need not be taken from experiment, since
it is predicted by the model. Thus, the experimental error on the model
test is reduced as compared to the model independent values shown in 
figure~\protect{\ref{vvppdependence.fig}}.
As mentioned above, the value in equation~(\ref{VVPPspincounting.eqn}) is
part of a
consistency check of a specific model and should not be interpreted as a
stand-alone measurement.

In principle, a comparison 
of the hadronisation fractions $\fPDsstp$ and $\fPDsp$ yields a
model independent measurement of $\VVPP$.
However, the result from section~\ref{dsstp.sec} cannot be interpreted 
in terms of vector and pseudoscalar meson production in charm hadronisation,
since no information on the separation of the
$\c\!\to\!\PDsstp$ and $\b\!\to\!\PDsstp$ components exists.

%        -------------------
\section{Summary and Conclusion}
%        -------------------
\label{conclusion.sec}

A first measurement of the hadronisation fraction $\fPDstz$ 
in $\PZz\!\to\!\cc$ decays is presented:
\begin{eqnarray*}
f(\c\!\to\!\PDstz) & = & 
\protect{\VfPDstz}
\,.
\end{eqnarray*}
This result is 
consistent with the expectation from isospin invariance.
The production rate of $\PDsstp$ mesons in hadronic $\PZz$ decays has been 
measured for the first time:
\begin{eqnarray*}
  \bar{n}(\PZz\!\to\!\PDsstp) \, Br(\PDsstp\!\to\!\PDsp\gamma)
  \, \protect{\BrDsp} \, \protect{\BrPhi} \hspace{3cm} & &
 \\
  \hspace{5cm} = 
  \protect{\VnbarPDsstpBrsummary}
 \,. & & 
\end{eqnarray*}

The relative production of vector 
charmed mesons in $\PZz\!\to\!\cc$ events, $\VVP$, is
evaluated both considering inclusive production ($\VVPF$) and taking
into account the effects of secondary production in $\PDstst$ decays
($\VVPP$). A value of 
\[
\VVPF = 
\protect{\VVVPF}
\]
has been derived from OPAL measurements, consistent with previous results.
The dependence of $\VVPP$ on the unmeasured $\PDstst$ multiplicities is
determined in a model independent calculation, where a weak dependence
on the production of the unmeasured broad $\PDstst$ resonances has been
found. 
From the determination of $\VVPP$, it is found that for the 
description of the production of charmed mesons in $\PZz\!\to\!\cc$ decays,
a simple spin counting picture is disfavoured by 
$\protect{\VnsigmaVVPP}$ 
standard deviations.

%        -------------------
\section*{Acknowledgements:}
%        -------------------

We particularly wish to thank the SL Division for the efficient operation
of the LEP accelerator at all energies
 and for
their continuing close cooperation with
our experimental group.  We thank our colleagues from CEA, DAPNIA/SPP,
CE-Saclay for their efforts over the years on the time-of-flight and trigger
systems which we continue to use.  In addition to the support staff at our own
institutions we are pleased to acknowledge the  \\
Department of Energy, USA, \\
National Science Foundation, USA, \\
Particle Physics and Astronomy Research Council, UK, \\
Natural Sciences and Engineering Research Council, Canada, \\
Israel Science Foundation, administered by the Israel
Academy of Science and Humanities, \\
Minerva Gesellschaft, \\
Benoziyo Center for High Energy Physics,\\
Japanese Ministry of Education, Science and Culture (the
Monbusho) and a grant under the Monbusho International
Science Research Program,\\
German Israeli Bi-national Science Foundation (GIF), \\
Bundesministerium f\"ur Bildung, Wissenschaft,
Forschung und Technologie, Germany, \\
National Research Council of Canada, \\
Research Corporation, USA,\\
Hungarian Foundation for Scientific Research, OTKA T-016660, 
T023793 and OTKA F-023259.\\

\newpage

\begin{appendix}
\section{Model Independent Calculation of 
         \boldmath\protect{$\VVPP$}\unboldmath}
In this appendix, details of the calculations that lead to 
the dependence of the model independent value of $\VVPP$ on the 
production of the broad $L$=1 charmed mesons (equation~(\ref{vvpi.eqn}))
are provided.

As discussed in the text, the contributions from decays of $\PDstst$
mesons have to be taken into account when calculating $\VVPP$.
Two assumptions are made on $\PDstst$ production:
\begin{list}{$\bullet$}{\itemindent=0pt \topsep=0pt
              \parskip=0pt \parsep=0pt \itemsep=0pt 
              \partopsep=0pt}
\item
Since $\PDststp$ mesons have not yet been observed in $\PZz\!\to\!\cc$
decays, isospin invariance is assumed to be valid which yields equal 
production rates for corresponding $\PDststz$ and $\PDststp$ spin states.
\item
For each light quark flavour q=u,d,s, two broad and two narrow $L$=1
${\rm c\bar{q}}$ mesons are predicted. For the ${\rm c\bar{u}}$ system, 
both narrow resonances ($\PDststCz$ and $\PDststDz$) have been measured
at OPAL, whereas only one narrow ${\rm c\bar{s}}$ resonance ($\PDsststCp$)
has been measured. Thus, the additional assumption is made that the
relative production of the different $L$=1 spin states does not depend on the
flavour of the light quark.
\end{list}
Under these two assumptions, the production of any $L$=1 charmed meson
can be expressed in terms of measured rates and two unknown parameters,
$\fPDststAz$ and $\fPDststBz$.

The OPAL ${\rm D_{s1}^+}$ production measurement~\cite{bib-Ddoublestar} 
assumes a $100\%$ branching
for ${\rm D_{s1}^+}\!\to\!{\rm D}^{(*)}\PK$ decays (\ie~to non-strange,
charmed non-orbitally excited mesons). Therefore, any contribution
from $\PDsststp\!\to\!{\rm D_{\s}^{(*)+}}{\rm X}$ 
decays is implicitly taken into account when using the measured
${\rm D_{s1}^+}$ multiplicity for an evaluation of $\VVPP$.  

The quantity $\VVPP$ is calculated as follows:
\begin{eqnarray}
  \protect{\VVPP} 
 & = &
  \frac{ V^{\rm prim} }{ V^{\rm prim} + P^{\rm prim} }
 \\
 & = &
  \frac{ V^{\rm eff} - \left( T_{\rm (s)}\!\to\! V \right) }
       { (V+P)^{\rm eff} - \left( T_{\rm (s)}\!\to\! V\ {\rm or}\ P \right) }
 \ ,
\end{eqnarray}
where in the above formula, $V$ and $P$ stand for non-strange charmed
vector and pseudoscalar mesons, respectively, whereas $T_{\rm (s)}$ denotes
the sum of non-strange and strange charmed tensor mesons ($\PDstst$ and 
$\PDsstst$).
Thus,
\begin{eqnarray}
  \protect{\VVPP}
 & = &
 \frac{   2 \fPDstp
        - 2 f(\c\!\to\!\PDststz\!\to\!\PDstzp)
        -   f(\c\!\to\!\PDsststp\!\to\!\PDstzp) }
      {     \fPDz + \fPDp
        - 2 \fPDststz
        -   \fPDsststp }
 \ ,
\end{eqnarray}
where the factors of 2 are introduced from the assumptions
$\fPDstz = \fPDstp$ and $\fPDststz = \fPDststp$. 

The symbol $f(\c\!\to\!\PDstst\!\to\!\PDstzp)$ is a shorthand notation
for the hadronisation fraction of a charmed quark to a $\PDstst$ meson
times the fraction of $\PDstst$ mesons decaying to a $\PDstzp$.
Spin and parity conservation restrict the possible decays of the
different $\PDstst$ spin states, as is illustrated in
table~\ref{Dststproperties.table}, where
a summary of predictions and measurements~\cite{bib-PDG96} 
for the four $\PDststz$ spin states is given.
Thus, the fraction of $\PDstst$ mesons decaying to a $\PDstzp$ depends
both on the relative production of the different $\PDstst$ spin states
and on the branching ratio $Br(\PDststD\!\to\!\PDstzp)$, since 
for the $\PDststD$ resonance, decays to both $\PDst$ and $\PD$ are allowed.
Due to phase-space effects, the branching fraction 
$Br_\s^{**} \equiv Br(\PDsststDp\!\to\!\PDstzp)$ of
$\PDsststDp$ mesons to a $\PDstz$ or $\PDstp$ is expected to differ
from the corresponding quantity $Br^{**} \equiv Br(\PDststDz\!\to\!\PDstzp)$ 
for non-strange $\PDstst$ mesons, see table~\ref{vvptable.table}.
\begin{table}[htb]
  \vspace{-10pt}
  \begin{center}
    \begin{tabular}{|l||c|c|c|c|}    
                          \hline
name  & ${\rm D_0^{*0}}$ & ${\rm D_1'^0}$ & ${\rm D_1^0}$ & ${\rm D_2^{*0}}$ \\
\hline
\hline
$j=s_{\rm q}+L$      & 
\multicolumn{2}{c|}{$(1/2)^+$} & \multicolumn{2}{c|}{$(3/2)^+$} \\
decay            & \multicolumn{2}{c|}{S-wave} & \multicolumn{2}{c|}{D-wave}\\
\cline{2-5}
spin-parity $J^{\rm P}$
                 &   $0^+$  &  $1^+$    &  $1^+$    &  $2^+$  \\
decay channels&$\PD\pi$&$\PDst\pi$&$\PDst\pi$&$\PD\pi$, $\PDst\pi$\\ 
width (MeV) \protect\cite{bib-PDG96} &$\sim 100$&$\sim 100$
                 & $18.9\pm ^{4.6}_{3.5}$ & $23\pm 5$ \\ 
mass (MeV) \protect\cite{bib-PDG96}   &            
      \multicolumn{2}{c|}{(not observed)}    &$2422.2\pm 1.8$& $2458.9\pm 2.0$ 
\\
 \hline
    \end{tabular}
  \end{center}
\caption{\label{Dststproperties.table}
     Properties of the neutral, non-strange 
     excited $\PD$ mesons. 
     For the decay channels, the same restrictions as shown here apply to 
     $\PDststp$ and 
     $\PDsststp$ mesons, where in the case of $\PDsststp$ decays,
     pions have to be replaced by kaons.} 
\end{table}

Using the shorthand notation $f_{\rm D_X} \equiv f(\c\!\to\!{\rm D_X})$, 
equation~(\ref{vvpi.eqn}) can then be derived as follows:
\begin{eqnarray}
\label{vvpp.eqn}
\protect{\VVPP}
& = & 
\frac{ 2 f_{\PDst}
     - 2 \left( f_{\PDststBz} + f_{\PDststCz} + Br^{**} f_{\PDststDz} 
         \right)
     - \frac{ f_{\PDsststCp} }{ f_{\PDststCz} }
       \left( f_{\PDststBz} + f_{\PDststCz} + Br_{\s}^{**} f_{\PDststDz} 
         \right)
     }
     { f_{\PDz} + f_{\PDp}
     - 2 \left( f_{\PDststAz} + f_{\PDststBz} + f_{\PDststCz} + f_{\PDststDz} 
         \right)
     - \frac{ f_{\PDsststCp} }{ f_{\PDststCz} }
       \left( f_{\PDststAz} + f_{\PDststBz} + f_{\PDststCz} 
            + f_{\PDststDz} \right)
     }
\\
\nonumber
 & =: & 
\frac{ A - R_\c f_{\PDststBz} }
     { B - R_\c \left( f_{\PDststBz} + f_{\PDststAz} \right) }
\,,
\end{eqnarray}
where in the transition to the second line,
both numerator and denominator are multiplied by a factor of 
$
R_\c /\!\left( 2 + \frac{ f(\c\to\PDsststp) }{ f(\c\to\PDststz) } \right)
$.
The parameters $A$ and $B$ are then given by
\begin{eqnarray}
\nonumber
 A 
 & = &
 R_\c \, \left[  2 \, f(\c\!\to\!\PDstzp)
               - 2 \left( f(\c\!\to\!\PDststCz) + f(\c\!\to\!\PDststDz\!\to\!\PDstzp) 
                   \right)
% the following \right. is to satisfy latex...
        \right.
\\&&\ \ \ \ \ \
        \left.
               - \left( f(\c\!\to\!\PDsststCp) + f(\c\!\to\!\PDsststDp\!\to\!\PDstzp) 
                 \right)
        \right]
 / \left[ 2 + \frac{ f(\c\!\to\!\PDsststp) }{ f(\c\!\to\!\PDststz) } \right]
\enspace{\rm and}
\\
\nonumber
 B
 & \equiv  &
 R_\c \, \left[   f(\c\!\to\!\PDz) + f(\c\!\to\!\PDp)
                - 2 \left( f(\c\!\to\!\PDststCz) + f(\c\!\to\!\PDststDz) \right)
        \right.
\\&&\ \ \ \ \ \
        \left.
                - \frac{ f(\c\!\to\!\PDsststp) }{ f(\c\!\to\!\PDststz) } 
                  \left( f(\c\!\to\!\PDsststCp) + f(\c\!\to\!\PDsststDp) \right) 
        \right]
 / \left[ 2 + \frac{ f(\c\!\to\!\PDsststp) }{ f(\c\!\to\!\PDststz) } \right]
\,,
\end{eqnarray}
respectively.
\end{appendix}

\newpage

\end{document}